\documentclass[aps,preprint,pre,showpacs,groupedaddress]{revtex4-1}
\usepackage{graphicx}

\begin{document}


\title{Domain Nucleation and Confinement In Agent Controlled Bistable Systems}


\author{Dorjsuren Battogtokh}
\email{dbattogt@vt.edu}
\thanks{}
\affiliation{ The Institute of Physics and Technology, Mongolian Academy of Sciences, Ulaanbaatar 51, Mongolia}
\affiliation{ Department of Biological Sciences, Virginia Polytechnic and State University, Blacksburg, Virginia 24061, USA}


\date{\today}

\begin{abstract}
We report a new mechanism of pattern formation in growing bistable systems coupled indirectly. A modified Fujita et. al. model is studied as an example of a reaction-diffusion system of nondiffusive activator and inhibitor molecules immersed in the medium of a fast diffusive agent.  Here we show  that, as the system grows, a new domain nucleates spontaneously in the area where the local level of the  agent becomes critical. Newly nucleated domains are stable and the pattern formation is different from Turing's mechanism in monostable systems. Domains are spatially confined by the agent even if the activator and inhibitor molecules diffuse. With the spatial extension of the system, a larger domain may undergo a wavenumber instability and the concentrations of active molecules within the neighboring elements of a domain can become sharply different. The new mechanism reported in this work can be generic for pattern formation systems involving multistability, growth, and indirect coupling.

\end{abstract}

\pacs{ 05.10.-a, 87.17.Aa, 87.18.Hf}

\keywords{}

\maketitle


Turing instability is the most well known mechanism of pattern formation in dissipative systems \cite{turing}, with the critical condition that the diffusion length of an inhibitor significantly exceeds the diffusion length of an activator \cite{mein}.  Under this condition, a periodic pattern emerges at a certain critical wavenumber near the stable uniform solution in a monostable system \cite{murray,asm}; in a bistable system,  periodic patterns can be developed near both of the stable uniform solutions \cite{bataa}, or between the bistable states, depending on initial conditions \cite{metens}. Turing patterns can be robust in growing systems \cite{maini} and on complex networks \cite{mikh}. 

Recently, Fujita et. al. proposed a mathematical model for pattern formation in  growing shoot apical meristem (SAM) \cite{fujita}. The emergence of new stem cell domains, where the concentration of the master protein $WUS$ is notably elevated compared to other zones of SAM,  is explained by the mechanism of Turing instability.  The authors  assumed  that both active variables, activator and inhibitor, are diffusive in the lateral directions, though there is no clear biological evidence about the diffusive nature of the proteins $WUS$ and $CLV1$ \cite{fujita,laux,meyer,meyer1,nikolaev,simon}. On the contrary, the experimental observation of sharp discontinuities of $WUS$'s level in adjacent cells \cite{meyer} suggests that the activator can be non-diffusive or slowly diffusive. A question arises as to whether domain confinement and new domain formation are possible in activator and inhibitor models, in particular in the WUS-CLV network, when the condition of Turing instability is not fulfilled. 

In this work we are concerned  with a mechanism of pattern formation in a class of reaction diffusion systems where  activator and inhibitor variables can be non-diffusive, but the coupling is carried out by  a fast diffusive variable. Such a system was introduced by Kuramoto for indirectly coupled biological cells, and it can be described by the model \cite{kura},
\begin{eqnarray}
\epsilon \frac{\partial  H}{\partial t}=- H +D_H \Delta_r H + \sum\limits_{j=1}^N w(\bf {X}_j) \delta (\bf{r-r}_j), \nonumber\\
\frac{\partial  {\bf X_i} }{\partial t}={\bf F}({\bf X}_i)+{\bf g}(H({\bf r}_i,t)), 
\end{eqnarray}
where, $H(x,t)$ is the diffusive variable, ${\bf X_i}$ represents the concentrations of chemical molecules in the cell $i$, $N$ is the number of cells, and ${\bf F}$'s are nonlinear functions. When $\epsilon <<1$, $H(r,t)$ is a fast variable whose dynamics are instantly dependent on a component of the vector of concentrations ${\bf X}$, through a function $w$.  Eq. (1) has been  studied when the functions ${\bf F}({\bf X})$ describe oscillatory dynamics \cite{kura,battphysA}. 

Let us use the mathematical model of shoot apical meristem by Fujita et. al. in our model of  an indirectly coupled system, in the case of a simple linear function for $\bf {g}$.  Here ``indirect coupling'' refers to the coupling of cells by diffusion of the dynamically inactive, byproduct variable $H$ \cite{kura}.

By  replacing ${\bf F}({\bf X})$ in Eq. (1) with the Fujita et. al. model, our model in spatial dimension one reads,
\begin{eqnarray}
\epsilon \frac{\partial  H}{\partial t}=-H + D_H \Delta_x H + \sum\limits_{j=1}^N X_j \delta (x-x_j), \nonumber\\
\frac{\partial  X_i}{\partial t}=\Phi(E+A_s X_i -B Y_i) -A_d X_i, \nonumber\\
\frac{\partial  Y_i}{\partial t}=C X_i -D Y_i + S_Y H(x_i), 
\end{eqnarray}
where the function $\Phi(Z)$ is given by the formula,
\begin{equation}
\Phi(Z)=\frac{A_d u_{max}}{2} (1+ \frac{\frac{2 Z}{A_d u_{max}}-1}{{\sqrt[n]{1+|\frac{2 Z}{A_d u_{max}}-1|^n}}}).
\end{equation}
In Eq. (2-3), $E$, $A_s$, $A_d$, $B$, $C$, $D$, $S_Y$, $u_{max}$, and $n$ are positive constants and $\Phi(Z)$ is a sigmoidal function  \cite{meyer1} with the values  in the range between $0$ and $A_d u_{max}$. $H(x_i)$ in Eq. (2) is the $H$'s value in the cell $i$. In the absence of $H$, Eq. (2-3) is the Fujita et. al.  model, where $X_i$ represents a hypothetical protein whose expression is controlled by WUS, and $Y_i$ represents the $CLV3$ protein in a given cell. 

Recent models of SAM suggest the existence of an unidentified diffusive factor in the regulation of SAM \cite{fujita,meyer1}. A diffusive factor  in a realistic model of SAM is termed as {\em stemness factor} \cite{simon}. We assume that this unidentified factor is a fast diffusive peptide-hormone,  which promotes the synthesis of local inhibitors by instantly sensing the activator concentration in the surrounding medium. The wiring diagram of our model is a combination of the wiring diagrams in Refs. \cite{nikolaev} and \cite{fujita}. It describes the interaction of the diffusive factor with the activator-inhibitor network of WUS and CLV (Appendix A).  In our model 
$X_i$ represents $WUS$, and  $Y_i$ represents the CLV1/CLV3 complex in a cell layer of SAM. We note that the form of the equation for $H$, and the linear coupling term, $ S_Y H(x_i)$, in the last equation of Eq. (2) have similarities with the corresponding equation and term in the mathematical model of Ref.  \cite{simon}. We first assume that both $X_i$ and $Y_i$ are non-diffusive, but the effects of slowly diffusing $X$ and $Y$ will be discussed later. 
 \begin{figure}
\vspace{-0.1in}
\includegraphics[width=0.7\linewidth,height=0.4\linewidth,scale=0.35]{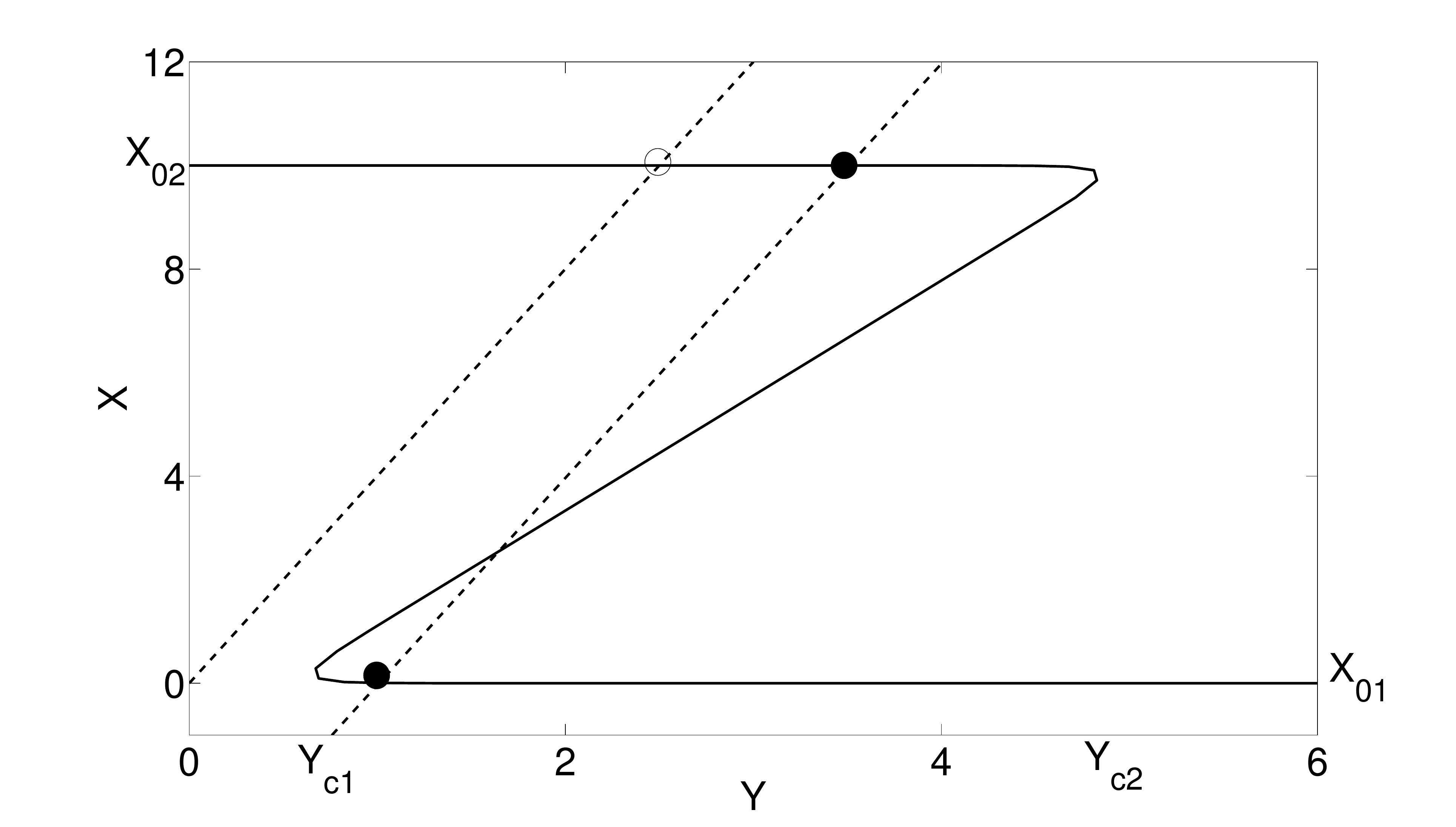} 
\caption{\label{fig1}  Nullclines of Eq. (5). Solid line, the nullcline of $X$. Dashed lines, nullclines of $Y$. At $HS=0$, there is a single stable steady solution shown by the open circle. At $HS=1$, bistable solutions are possible, filled circles. The parameters are: $E=1$, $B=2$, $C=0.25$, $D=1$, $A_d=1$, $A_s=1.9$, $u_{max}=10$, $n=20$, $HS_0=0$, and $S_Y=1$. }
\end{figure}

By assuming a sufficiently dense and uniform distribution of cells \cite {kura}, the continuity limit of Eq. (2) can be taken, $X = \frac{X_i}{\delta x} $ and $Y = \frac{Y_i}{\delta x} $, for a small length increment $\delta x \rightarrow 0$. Then for $\epsilon<<1$, $H$ can be expressed by, 
\begin{equation}
\hat H(x,t) \approx {\frac{1}{2\sqrt{D_H}}} \int_0^L \mathrm{e}^{-\frac{ |x-x'|}{\sqrt{D_H}}}X(x',t)\,\mathrm{d}x',
\end{equation}
where $L$ is the system size.  Introducing  $HS=HS_0+S_Y \cdot \hat H(x,t)$, Eq. (2) is approximated by,
\begin{eqnarray}
\frac{\partial  X}{\partial t}=\Phi(E+A_s X -B Y) -A_d X, \nonumber\\
\frac{\partial  Y}{\partial t}= C X -D Y +HS.
\end{eqnarray}
In the limit $\sqrt{D_H} >> L$, $\hat H(x,t)$ can be replaced by the global coupling function of $X$, $\hat H_g \approx {\overline X}$. Let us assume that for certain initial conditions,  ${\overline X}$ can evolve into a fixed nonzero value. Then  the bistable solutions of Eq. (5), corresponding to the fixed value of ${\overline X}$, can be found from the intersections of the nullclines with a constant intercept $HS$, Fig. 1. The saddle-node points for the transition between monostable and bistable solutions  at $C=0$ are marked as $Y_{c1}$ and $Y_{c2}$ in Fig. \ref{fig1}. The lines crossing these points have  the intercepts $HS_{c1}  \approx D \cdot Y_{c1}-C \cdot X_{c1} $ and $HS_{c2}  \approx D \cdot Y_{c2}-C \cdot X_{c2}$, where $X_{c1}$ and $X_{c2}$ can be computed from the maximum and minimum of the $X$ nullcline.

We use the term domain for an area in a bistable system, where the local values of a variable are continuously higher or lower than its values in other areas of the system. In an extended bistable system, pattern formation depends on initial conditions. A linear stability analysis of Eq. (5)  (Appendix C) shows that a domain nucleation is possible in Eq. (5) near the saddle-node points, from small initial perturbations of uniform states, $(X_{01},Y_{01})$ or $(X_{02},Y_{02})$. Since $X$ and $Y$ are nondiffusive, the maximum growth rate of the linearized system corresponds to large wavenumbers \cite{kura}. 

In Eq. (5) a domain nucleation is possible from certain initial distributions of $H$ even when $HS_0=0$.  Let us consider initial conditions such that the size of the initial $X_{01}$ domain is $\Delta X_{01}$, and the size of the  $X_{02}$ domain is  $\Delta X_{02}$. Then a stationary  two-domain solution can exist, with the global value $H_g \approx \overline X_0 =\frac{\Delta X_{01}}{L} \cdot X_{01}+\frac{\Delta X_{02}}{L} \cdot X_{02}$. For $L=1$ and $X_{01} \approx 0$, the critical sizes  for the existence of the two-domain solution are $\Delta_{X_{max}}=1-\frac{ HS_{c1}}{S_Y X_{02}}$ and $\Delta_{X_{min}}=1-\frac{HS_{c2}}{S_Y X_{02}}$.

\begin{figure}
\vspace{-0.1in}
\includegraphics[width=1.\linewidth,height=0.4\linewidth,scale=0.4]{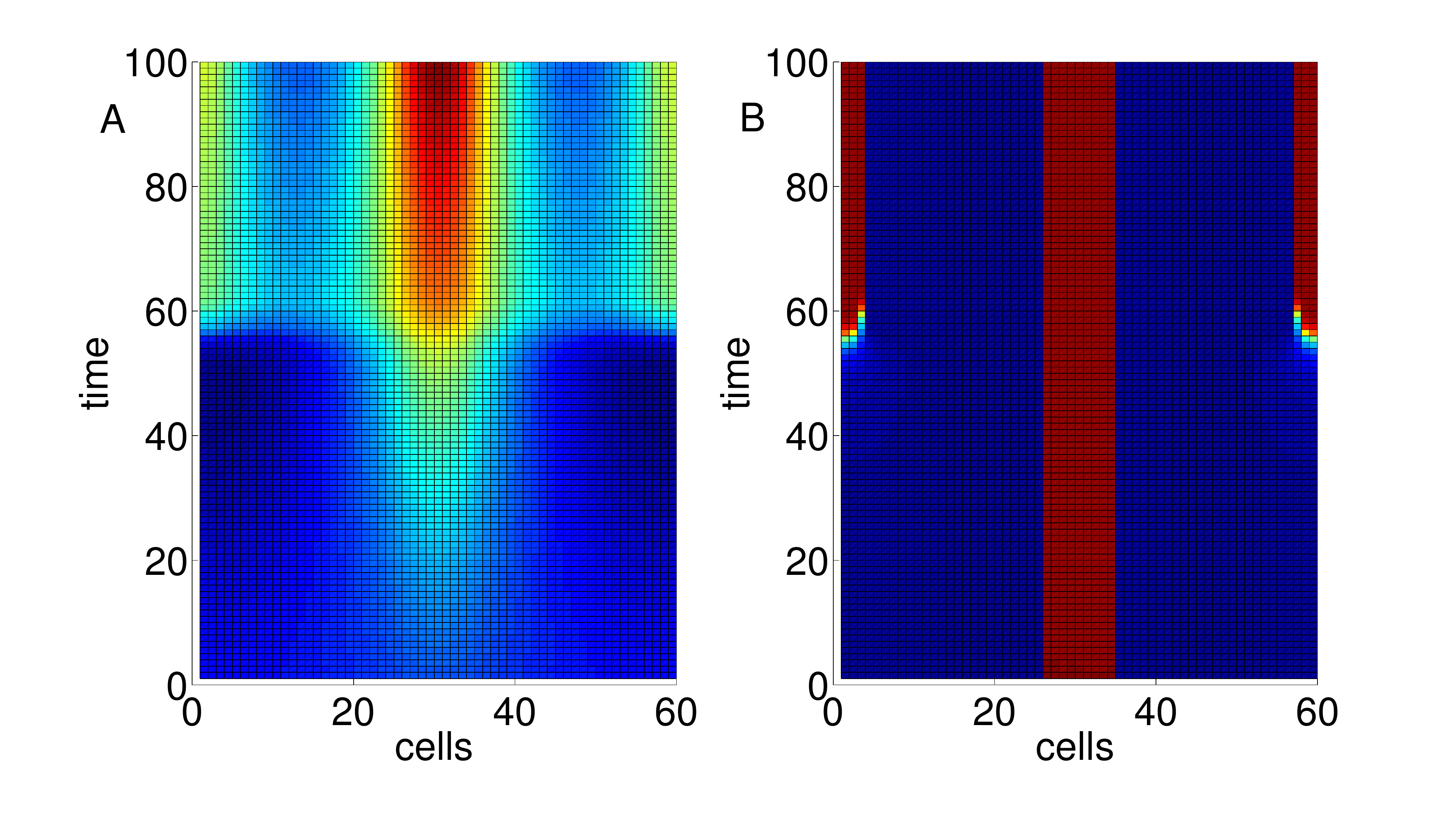} 
\caption{\label{fig2} (Color online) Domain nucleation in a growing system described by Eqs. (2-3). Here we assume  that the number of cells is fixed but the distances between the cells are increasing at a constant rate. A) Space-time dynamics of $H$. Higher values of $H$ are shown in red(white), lower values are shown in dark blue(black). B) Space-time dynamics of $X_i$'s. Higher values (near $X_{02}$) are shown in red(white), lower values (near $X_{01}$) are shown in dark blue(black). Parameters are the same as in Fig. 1, except $D_H=100$, $\epsilon=0.01$, $S_Y=0.5$,  $\Delta X_{02}=1.8$, and $L_{S_0}=12$.}
\end{figure}

\begin{figure}
\vspace{-0.1in}
\includegraphics[width=1.\linewidth,height=0.4\linewidth,scale=0.4]{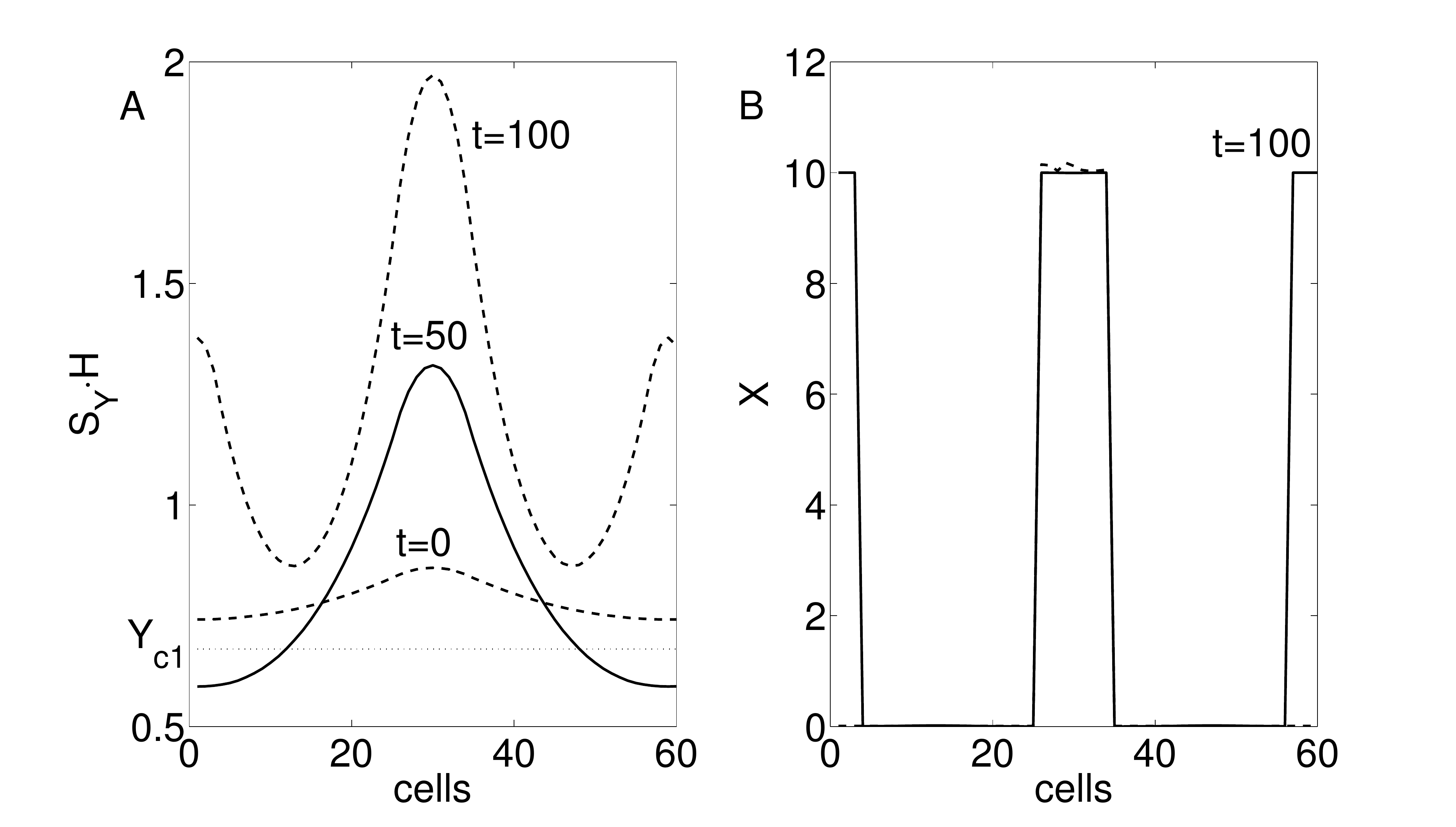} 
\caption{\label{fig3} Snapshots of $H$ and $X$ from Fig. 2 at different time moments. A) Distributions of $H$. B)  Distributions of $X_i$'s.}
\end{figure}

At the global coupling limit, the two domain solution, which can be formed from the initial conditions described above, is stable in Eq. (5), in the interval $H_{c1} \leq HS \leq H_{c2}$.  We seek to illustrate what happens when the coupling range is reduced from global to nonlocal, i.e., $\Delta_{X_{min}}<<\sqrt{D_H} < L$. In this limit, $HS(x,t)$ can be nonuniform for a two-domain solution in Eq. (5), such that it will be higher near  the center of the upper domain $X_{02}$, but lower in the areas further away from the center. With further reduction of the coupling range, it may become possible  that in some areas $HS(x,t)<Y_{c1}$ ($HS_{c1} \approx Y_{c1}$, for $D=1$) but in the bulk of the system $HS(x_{bulk},t)>Y_{c1}$. In other words,  a domain with the value $X_{02}$ may be nucleated in the area where $HS(x,t)<Y_{c1}$, because $X_{02}$ is the only stable solution below $Y_{c1}$ in Fig. \ref{fig1}. After the nucleation,  $H(x,t)$ will be quickly adjusted and  $HS(x,t)>HS_{c1}$ everywhere. If the intersection of the nullclines is near the point $Y_{c2}$ in Fig. 1, the nucleating domain is $X_{01}$ and consequently $HS(x,t)<HS_{c2}$ everywhere. 

To study the domain nucleation associated  with the reduction of the coupling range from global to  nonlocal, we simulated Eq. (2-3). For detailed numerical simulations of Eq. (2-3),  the method proposed in Ref. \cite{kura1} is more suitable, because when the system size is small and $\epsilon << 1$, the finite-difference scheme is stable only at a small time step. We simulated Eq. (2-3) with periodic and no-flux boundary conditions. The initial distributions of $X_i$'s are chosen such that the $X_{02}$ domain is in the center of the system. For $Y_i$ and $H$, uniform initial distributions are chosen. The effects caused by the increase of the system length can be simulated in Eq. (5)  by increasing the ratio, $\frac{L}{\Delta X_{02}}$, at fixed $L$ (see Fig. 4). But, to make our simulations of Eqs. (2-3) relevant to modeling of SAM growth, we first let the length of the simulation domain, $L_S$, increase with time. In our simulations, $L_S$  is increased in the time interval $t <50$ as $L_S=L_{S_0} (1+\frac{t}{25})$, where $L_{S_0}=12$ is the initial size. After $t \geq 50$ the system size was fixed.  Fig. 2 shows a space-time plot of the simulations, and Fig. 3 shows spatial distributions of $H$ and $X_i$ at different time moments. As time increases, the distribution of $H(x,t)$ gets higher in the center of the system, but lower in the areas further away from the center of the system, Fig. 2A.  In the areas where the distribution of $S_Y \cdot H(x,t)$ becomes less than the critical value $Y_{c1}$,  new domains are nucleated, Fig. 3A.  We obtained similar results as in Fig. 2, when the length of the system was increased by adding new elements at the boundaries, in the case of no-flux boundary conditions.

\begin{figure}
\vspace{-0.in}
\includegraphics[width=1.\linewidth,height=0.4\linewidth,scale=0.65]{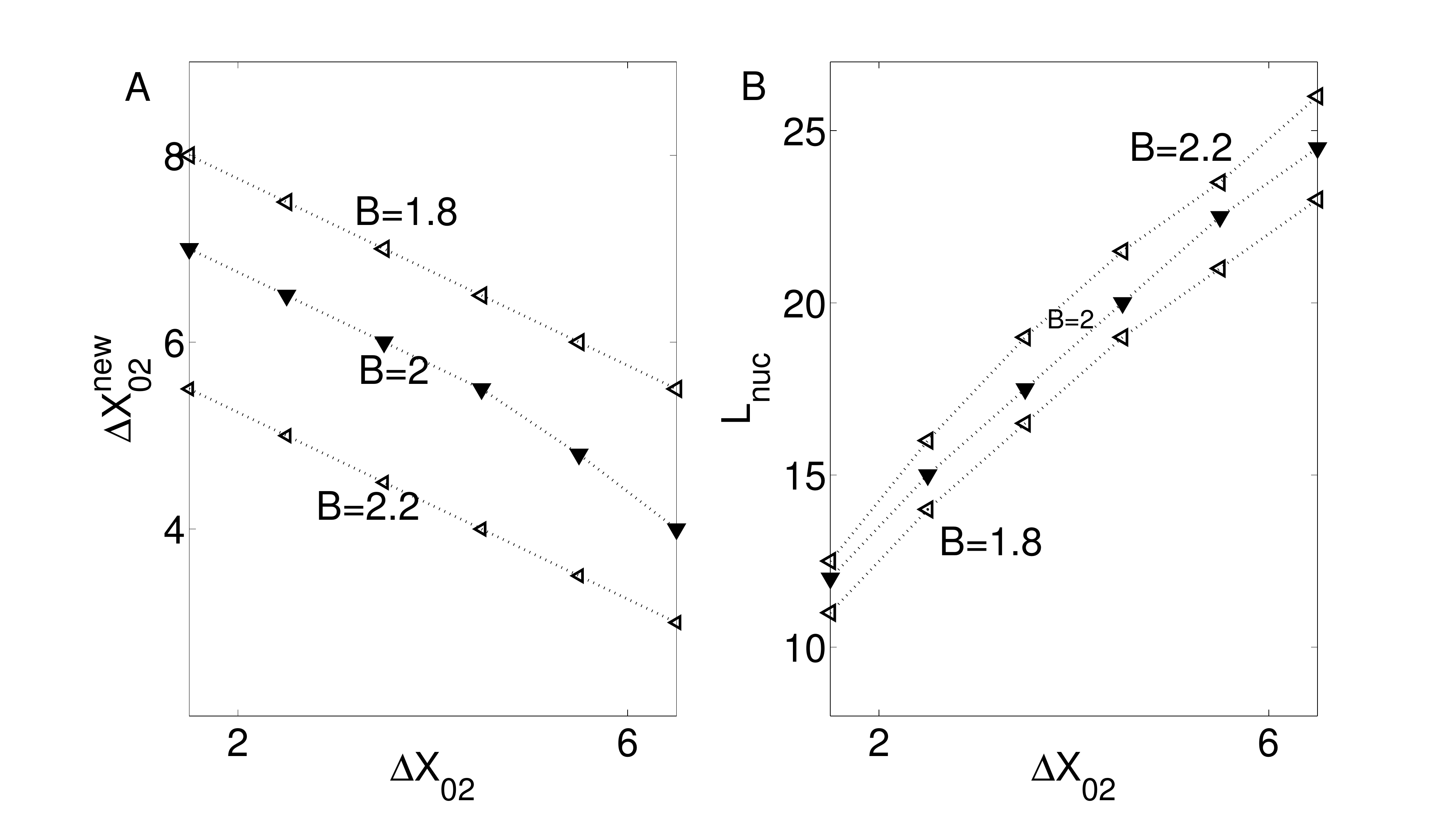} 
\caption{\label{fig4} Domain nucleation at different values of the size of initial domain. A) The size of newly nucleated domain, $\Delta X_{02}^{new}$,  vs the size of initial domain $\Delta X_{02}$. B) The distance from the center to the new domain, $L_{nuc}$, vs the size of initial domain $\Delta X_{02}$. Other parameters are the same as in Fig. 2. The size of the system is fixed at $L_S=64$. }
\end{figure}

The distance from the center of the system to the location where a nucleation of the domain takes place is dependent on the size of the initial domain in the center of the system.  Assuming the size of the upper domain is $\Delta X_{02}$,  an estimate can be made from Eq. (4) for the distance from the initial domain in the center to the location of the new domain,
\begin{equation}
L_{nuc}=\sqrt{D_H} \mathrm{Ln}(\frac{ 2(X_{02} - X_{01}-X_{02} {\mathrm{e}}^{(\frac{-\Delta X_{02}}{{\mathrm{\sqrt{{D_H}}}}})})}{HS_{c1}-2 X_{01}}),
\end{equation}
where  $HS_{c1} \sim Y_{c1}$ is defined by the parameters of the model. Numerical simulations are in  qualitative agreement with Eq. (6) that the locations of new domains depend on the size of the initial domain $\Delta X_{02}$. In Fig. 4 we plot the results obtained from numerical simulations of Eq. (2-3) at different values of $B$. The size of the new domain decreases with the increase of $\Delta X_{02}$, whereas, the distance from the center to the location of the nucleation increases with the increase of $\Delta X_{02}$. Fig. 4 implies that unlike the Turing patterns in monostable systems, domains in Eq. (2-3) can have different sizes and their distances to each other can differ.  

\begin{figure}
\vspace{-0.in}
\includegraphics[width=1.\linewidth,height=0.4\linewidth,scale=0.5]{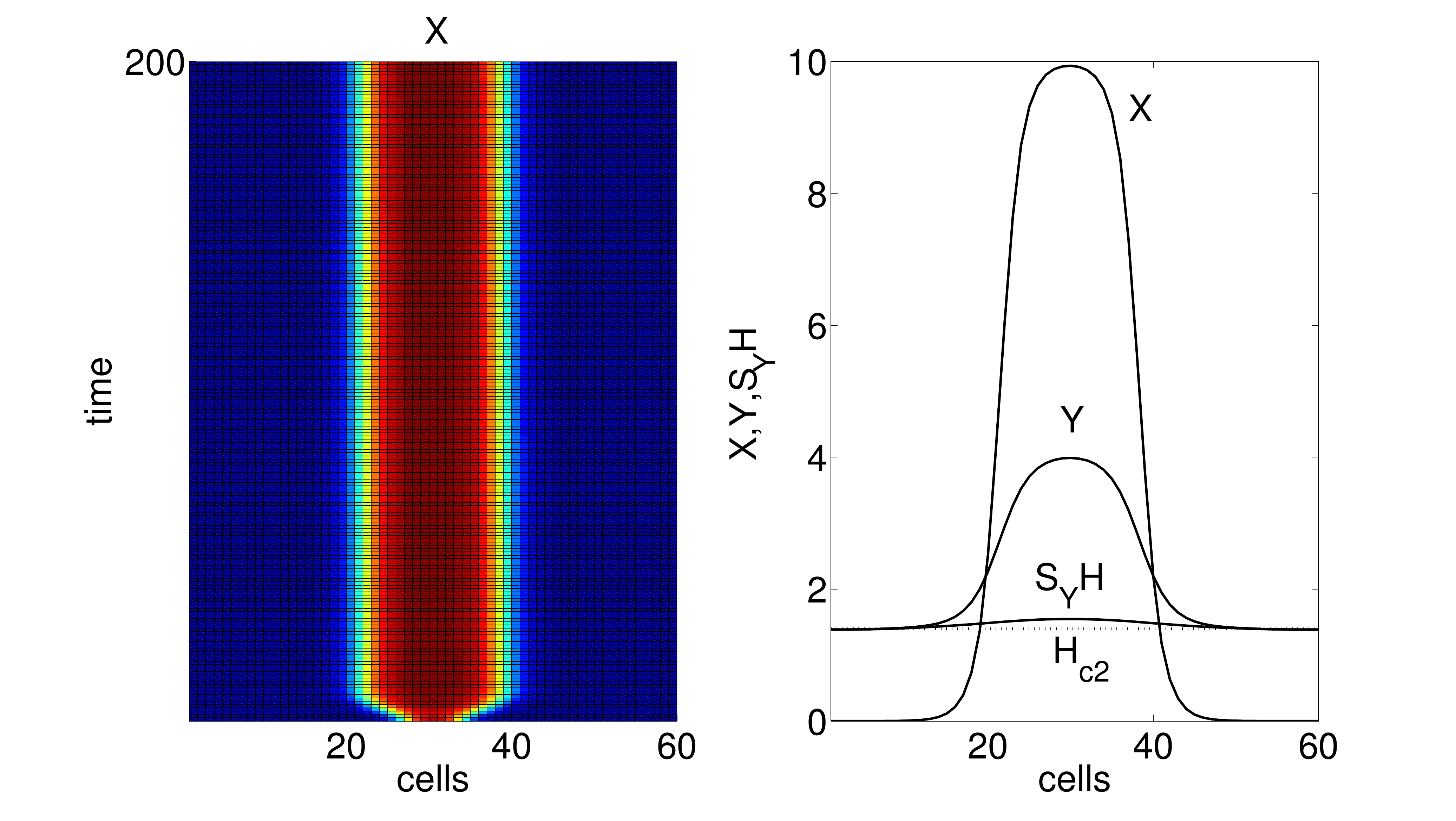} 
\caption{\label{fig5} (Color online) Domain confinement.  A) Space-time plot of $X_i$'s. B) Snapshots of $X_i$'s, $Y_i$'s, and $H$ at time $t=200$. Parameters are the same as in Fig. 2, except $D_X=0.1$, $D_Y=0.1$, and the system size is fixed at $S_L=12$.}
\end{figure}

A natural question then is whether the domain nucleation in Fig. \ref{fig2} is possible if $X$ and $Y$ are diffusive. We studied the effects of small diffusion($D_{X,Y}<0.3$),  by  adding the term $Dif_X=D_X (X_{i-1}- 2 X_{i}+X_{i+1})$ into the activator equation, and $Dif_Y=D_Y (Y_{i-1}- 2 Y_{i}+Y_{i+1})$ into the inhibitor equation of Eq. (2). Numerical simulations of Eq. (2-3) with the diffusion terms $Dif_X$ and $Dif_Y$ indicate that the domain nucleation is persistent for slowly diffusing activator and inhibitor. This is because a domain can be confined in our agent controlled system. Fig. 5 shows the  confinement of a domain when  $X$ and $Y$ are diffusive. The domain confinement in our system is the mutual  equilibrium of the upper and lower domains, controlled by the agent (Appendix B).  This confinement allows nucleation of new domains in growing systems, when $X$ and $Y$ slowly diffuse.

Simulations show that at a stronger coupling strength, a large $X_{02}$ ($X_{01}$) domain loses its stability, and $X_{01}$ ($X_{02}$) states are spontaneously generated. Fig. 6 shows stationary patterns obtained from  the simulations with random initial conditions near $X_i \approx X_{02}$ and long-wave  distributions of $Y_i$'s and $H$. The parameters are chosen such that the homogenous steady state is near the point $Y_{c2}$ in Fig. 1, and it is unstable to nonuniform fluctuations (Appendix C). In the absence of $X_i$ and $Y_i$ diffusion, the profile of $X_i$'s distribution resemble the  {\em chimera state} in nonlocally coupled oscillators \cite{chimera}, Fig 6A.  However, if $X_i$ and $Y_i$ diffuse slowly, and  $D_X<D_Y$, the pattern is smooth, but large jumps of the activator concentrations between the neighboring cells are possible, Fig. 6B.  Interestingly, in the experiments, $WUS$ expression in adjacent cells can be sharply different \cite{meyer}. 

When $D_H \sim D_{X,Y}$ and the system is in the monostable state near the saddle-node point $Y_{c1}$, depending on initial conditions, small amplitude Turing patterns can emerge in Eqs. (2-3) via the critical mode selection \cite{metens}. With the increase of $D_H$, the amplitude of these patterns may increase until its maximum and minimum reach the values of the bistable states, such that the pattern behaves like periodic domains in a bistable region (Appendix D). For two-variable bistable systems, these regular patterns are possible at $1 \leq \frac{D_Y}{D_X} \leq 6$, unlike for the monostable systems where the ratio is typically larger than 6 for pattern formation to be possible. Therefore, the emergence of regular shaped patterns in the bistable region of Eq. (2-3) can be explained by the pattern selection mechanism, as the continuation of the Turing patterns of a homogeneous steady state near the saddle-node points \cite{metens}. On the contrary, the size, location, and spatiotemporal dynamics of  not only regular patterns, but also of irregular patterns of Eqs. (2-3), in a wide range of the parameters and initial conditions, can be explained by the mechanism we described in Fig. 2. 

\begin{figure}
\vspace{-0.in}
\includegraphics[width=1.0\linewidth,height=0.35\linewidth,scale=0.5]{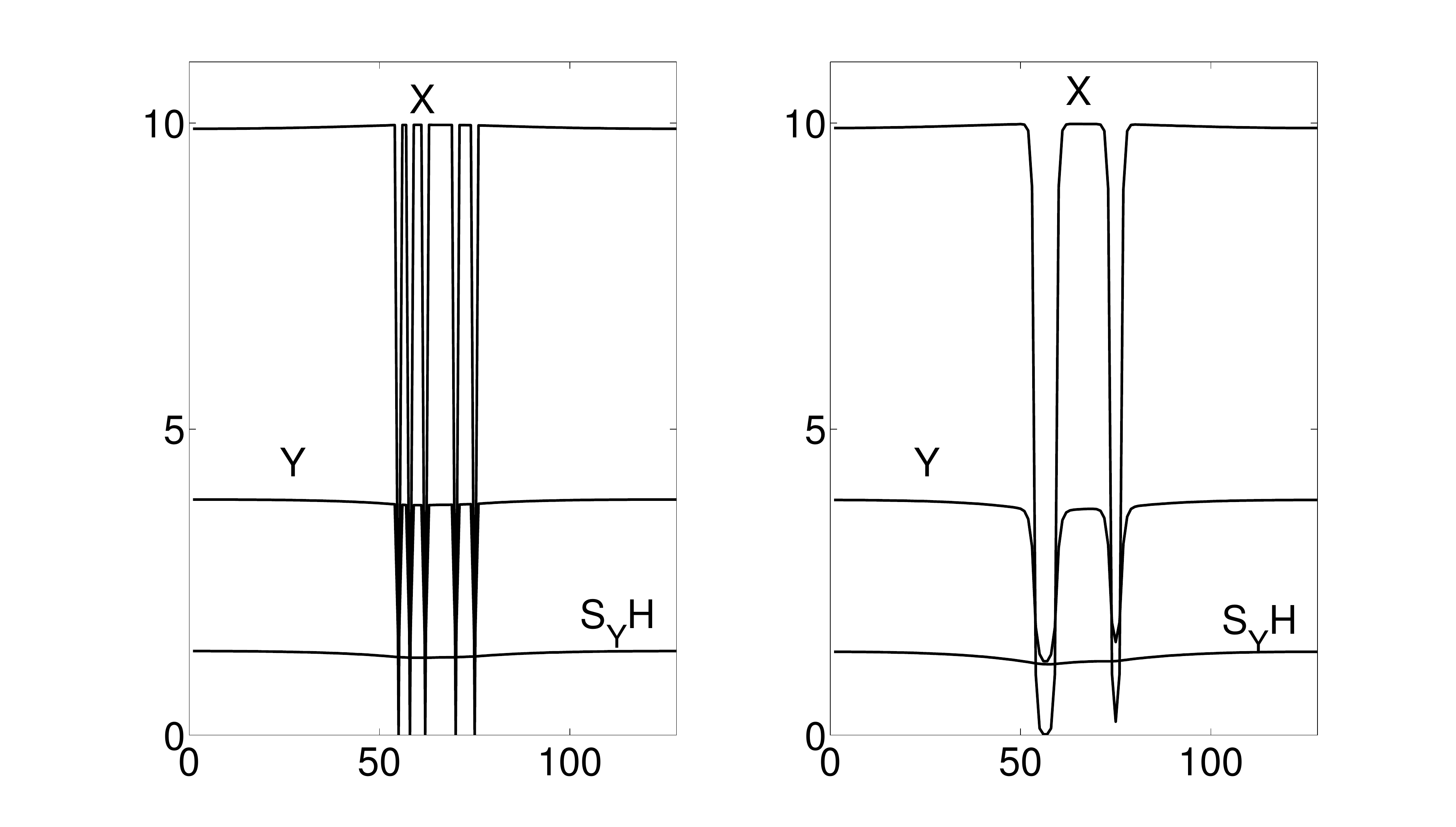} 
\caption{\label{fig6} A domain instability leading to the coexistence of large and small domain solutions. A) Nondiffusive $X$ and $Y$, B) $DX=0.035$ and $D_Y=0.1$. Parameters are the same as in Fig. 4, except $S_Y=0.14$,  and $B=2.5$}
\end{figure}

In summary, we studied domain formation and instability in growing bistable systems with a reaction diffusion model, where active variables are non-diffusive but immersed in a medium of a fast diffusive agent. We explained domain nucleation in such a system with a new mechanism. In contrary to the existing theory that explains pattern formation in bistable systems with the Turing mechanism of  {\em nascent} bistability \cite{metens}, the new mechanism explains it by the intrinsic transitions between coexisting states, controlled by the agent. The new mechanism offers alternative interpretation of  existing data and design of next experiments. The experimental data on SAM can be explained in terms of domain nucleation and front bifurcation, not by the critical mode selection of the Turing mechanism. Finally, we believe that the agent controlled pattern formation is generic for developmental biology, involving multistability, growth, and indirect coupling.

\appendix
\section{Wiring diagrams and minimal models of SAM}
\begin{figure}
\includegraphics[width=0.9\linewidth,height=0.35\linewidth,scale=0.35]{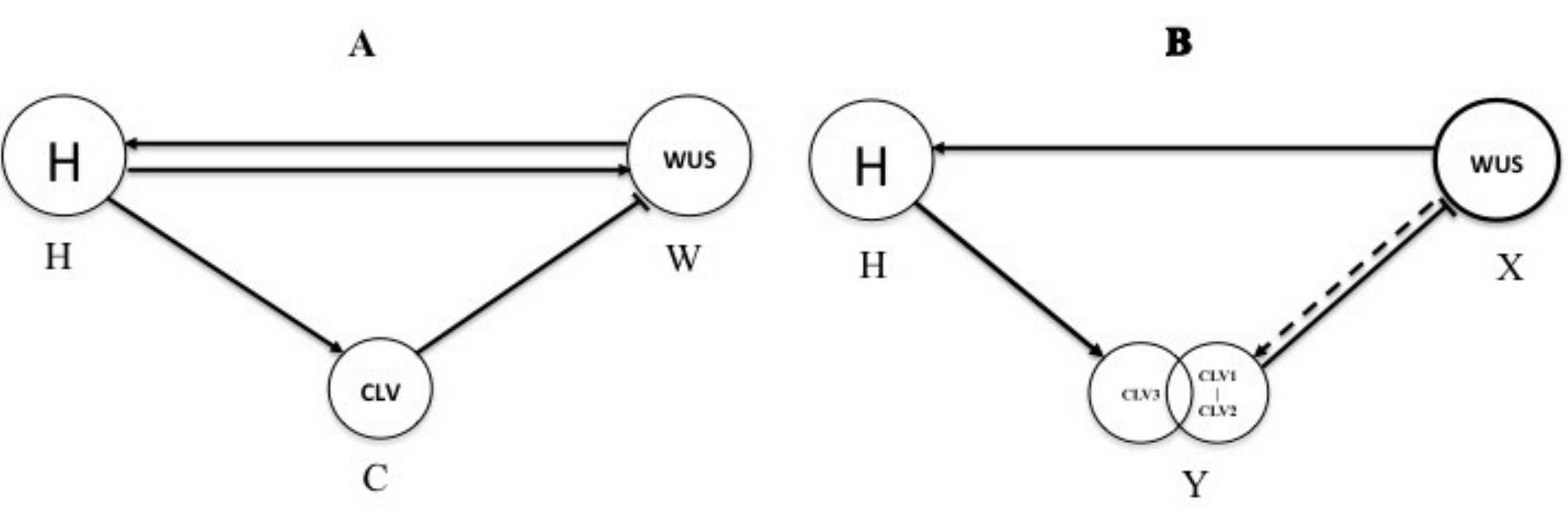} 
\caption{\label{Fig3S} Wiring diagrams of minimal models of SAM regulation. A) A diagram of a minimal model of SAM regulation by Nikolaev et al \cite{nikolaev}. B) A modified version of the diagram.  By the intense lines for $WUS$, we express a self-enhancement mechanism of $WUS$. The dashed line shows an activation of $Y$ by $WUS$. The last two features, the characteristics of activator-inhibitor interactions, are adopted from the Fujita et al model \cite{fujita}.}
\end{figure}

To explore the core mechanisms of SAM regulation, Nikolaev et. al. proposed a minimal mathematical model of SAM \cite{nikolaev}. The wiring diagram of the model is shown  in Fig.  \ref{Fig3S}. The model describes the interactions between $WUS$, $CLV$, and an unidentified factor $H$, in one dimensional model of a vertical section of SAM. It is given by the following ODE's,

\begin{eqnarray}
\frac{\partial  H_i}{\partial t} & = &  -d_h H_i +(H_{i-1}+H_{i+1}-2 H_i),  \nonumber\\
\frac{\partial  C_i}{\partial t} & = &  -C_i +g_C(h_C+T_{CH} H_i),\nonumber\\
\frac{\partial  W_i}{\partial t} & = &  -W_i +D_W (W_{i-1}+W_{i+1}-2 W_i) + g_W(h_w+T_{WH} H_i+T_{WC} C_i), 
\end{eqnarray}
where the cell index $i$ goes from $i=2,3,..N-1$, where $N$ is the number of cells. The boundary cells are described by the following set of equations, 
\begin{eqnarray}
\frac{\partial  H_1}{\partial t} & = &  -d_h H_1 +(H_2-H_1) + g_H(h_H+T_{HW} W_1),  \nonumber\\
\frac{\partial  H_N}{\partial t} & = &  -d_h H_N +(H_{N-1}-H_N),  \nonumber\\
\frac{\partial  W_1}{\partial t} & = &  -W_1 +D_W (W_2-W_1) + g_W(h_w+T_{WH} H_1+T_{WC} C_1),  \nonumber\\
\frac{\partial  W_N}{\partial t} & = &  -W_N +D_W (W_{N-1}-W_N) + g_W (h_w+T_{WH} H_N+T_{WC} C_N),  \nonumber\\
\frac{\partial  C_i}{\partial t} & = &  -C_i +g_C(h_C+T_{CH} H_i), i=1 \& N.  
\end{eqnarray}

\begin{figure}
\includegraphics[width=1\linewidth,height=0.5\linewidth,scale=1]{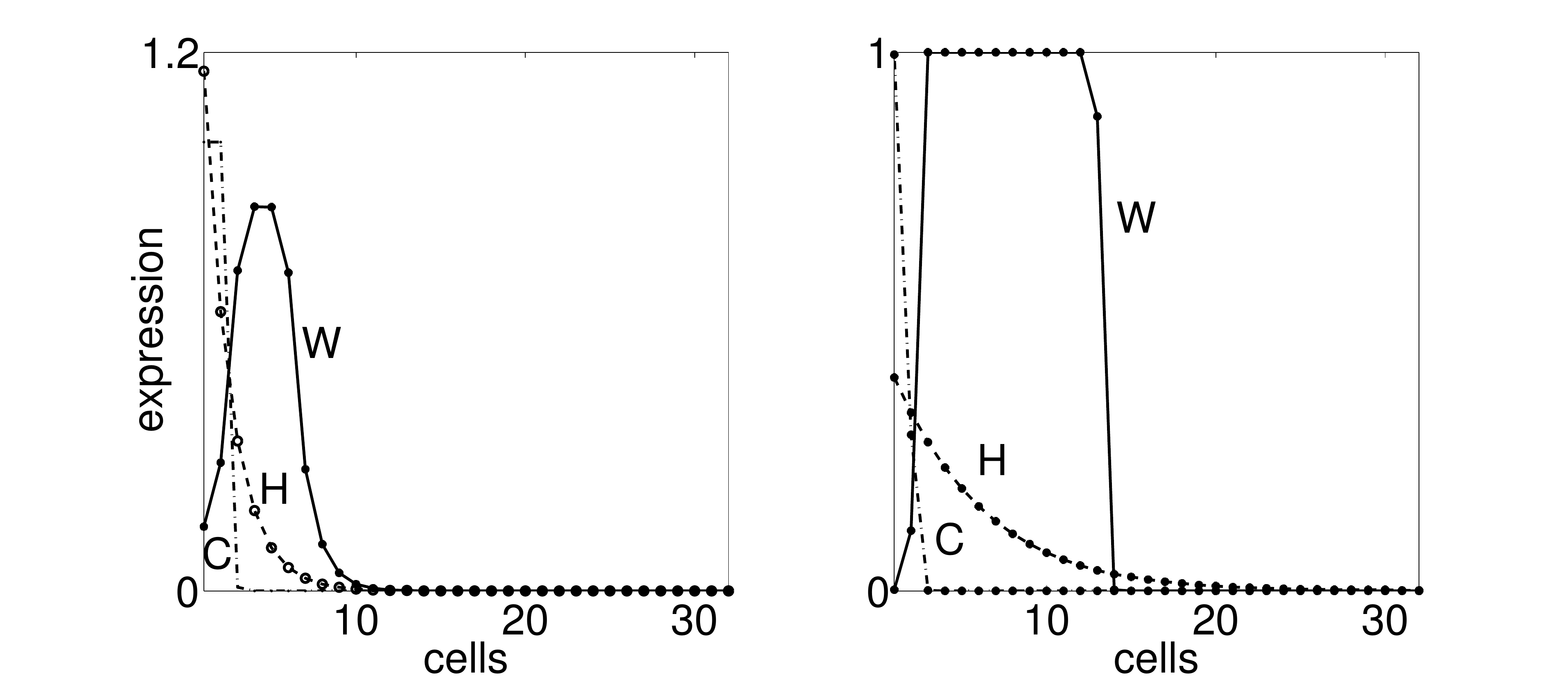} 
\caption{\label{FigNiko} Simulations of Eqs. (A1-A2). Left: $D_W=1$. Right: $D_W=0$. Other parameters are: $h_Y=-0.1$, $h_W=h_C=-88.9379$, $d_h=0.4$, $T_{WH}=2500$, $T_{WC}$=-2900, $T_{HW}=300$, and  $T_{CH}=250$.}
\end{figure}

In Eqs. (A1-A2), $C$ variable is non-diffusive. The function $g$ describes the interactions between the genes/proteins in Fig. \ref{Fig3S}, and it is given by a sigmoidal function, 
\begin{equation}
g(\xi)=\frac{1}{2}(1+\frac{\xi}{\sqrt{1+\xi^2}}).
\end{equation}
For more detailed descriptions and simulations of the model, Eq. (A1-A2), we refer to the Ref. \cite{nikolaev}. Here we simulated Eqs. (A1-A2) to show that the model displays dynamics similar to what one would expect from  bistable reaction-diffusion systems. Fig. \ref{FigNiko} shows stationary distributions of  $W$, $C$, and $H$ on a cell line of 32 cells. The distribution of $W$ in the stem cell zone of  Fig. \ref{FigNiko} is reminiscent of a domain in bistable systems, especially when $D_W=0$, Fig. \ref{FigNiko} right plot.   Fig.  \ref{FigNiko}  suggests that the argument of the function $g$, which is a linear combination of the levels of $W$, $C$ and $H$  fields, can switch the system between bistable states. Therefore, a question arises as to whether the closed forms of the models of SAM with nonlinear functions $g$ ($\Phi$ in Eqs. (1-2)) can display an intrinsic bistability.  And if so, what are the mechanisms of domain nucleation  and domain confinement in the bistable regime? What is the role of bistability in the models of SAM, in particular, in  the activator-inhibitor model of Fujita et al?

To answer these questions we modified the wiring diagram in Fig.  \ref{Fig3S} A to the one shown in Fig.  \ref{Fig3S} B.  We have chosen Fujita et al model because it displays Turing patterns and intrinsic bistability. Following the Fujita et al model, we assume that $WUS$ and $Y$ can be an activator-inhibitor system, where $WUS$ is self-enhancing, and also activating its inhibitor $Y$.  This way we consider the system in the modeling framework of Kuramoto \cite {kura}, as a reaction-diffusion system, coupled through an indirect, fast diffusive-field, $H$. 

In Eqs. (1-3) of the main text, the nonlinear function $g_X$ is replaced by $\Phi$,
\begin{equation}
g_X(\xi)=\Phi (E+T_{XH} H+A_s X +T_{XY} Y),
\end{equation}
where, we assume that $T_{XH}<<1$, and $T_{XY}=-B$.  Note that $A_d=1$ in Eq. (2). $g_Y$ is replaced by, 
\begin{equation}
g_Y (HS_0+T_{XY} X+T_{YH} H) \rightarrow HS_0+C X+S_{Y} H,
\end{equation}
where, following Fujita et al \cite{fujita}, we approximate the sigmoidal function $g_Y$ with the linear terms only. Note that $D=1$ in Eq. (2) of the main text. $g_H$ in our model is replaced by,
\begin{equation}
g_H (h_H+T_{HX} X) \rightarrow X,
\end{equation}
where $h_H=0$. We assume $H$ is a fast variable. As the goal of our model is to study the mechanisms of bistability, domain formation, and domain confinement, we study our model in a closed form. 

\section{Domain Confinement}
\subsection{Domain Potential}
\begin{figure}
\includegraphics[width=0.6\linewidth,height=0.4\linewidth,scale=0.5]{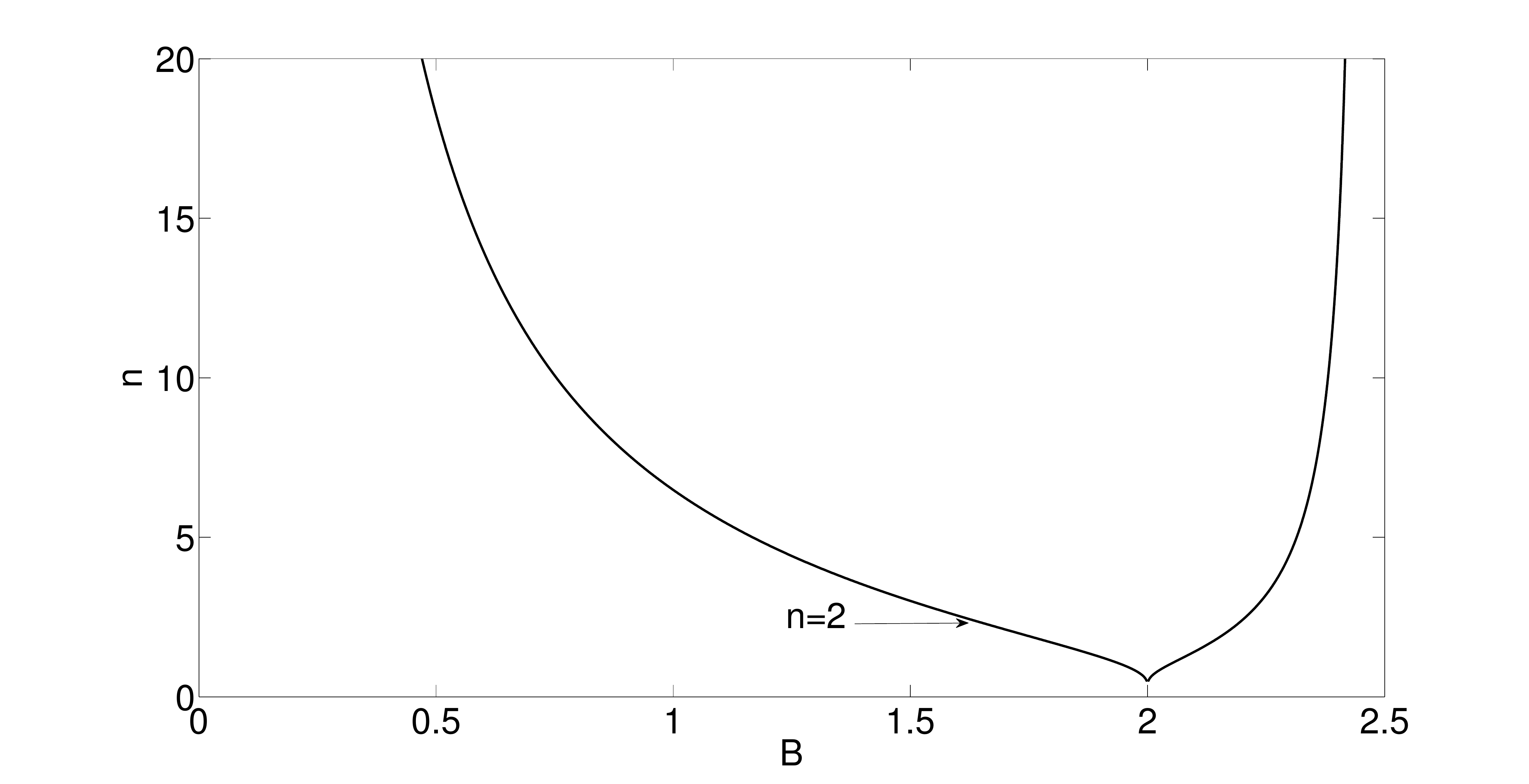} 
\caption{\label{Fig4S} A cusp bifurcation in Eq. (5) of the main text. Parameters are $A_s=1.8$, $A_d=1$, $B=2$, $C=0.25$, $D=1$, $E=0.1$, $u_{max}=10$, and $HS=0.8$. }
\end{figure}

To simplify our analysis, here we study the case of $n=2$ for Eq. (3) in the main text. Fig.  \ref{Fig4S} shows a cusp bifurcation in Eq. (5) of the main text, obtained from the continuation of the saddle-node bifurcation points in Fig. 1 of the main text, using $n$ and $B$ as the principal bifurcation parameters. Fig.  \ref{Fig4S} shows that at $n=2$, the bistability is still  present, although at $n=2$, the bistable region is narrow compared to the case when $n>2$. 

For further simplification, we next decouple the first equation in Eq. (5) of the main text  from the second equation, by assuming a constant $Y$, $Y=A_0$. Hence, in the case of a diffusive $X$, we obtain a single  PDE, 
\begin{eqnarray}
\frac{\partial  X}{\partial t} & = & f(X,A_0) +\Delta X, \nonumber\\
f(X,A_0) &=& \frac{A_d u_{max}}{2} + \frac{{(E+A_s X-B A_0)-0.5 A_d u_{max}}}{{\sqrt{1+(\frac{2 (E+A_s X-B A_0)}{A_d u_{max}}-1)^2}}} - A_d X.
\end{eqnarray}
Eq. (B1) has two stable solutions, $X_{01}$ and $X_{02}$, which can be obtained numerically by solving the equation  $f(X,A_0)=0$ at different $A_{0}$. From these solutions a table of bistable solutions at different values of $A_0$ can be built. 

The two stable solutions are connected by a front due to the presence of the diffusion term in Eq. (B1). Our goal is to find the condition when the front solution is motionless,  i.e. $v=0$, where $v$ is front velocity, depending on the model parameters.  By considering the nonlinear term $f(X,A_0)$ as the forcing term and the diffusion term as the dissipation term, we express $f(X,A_0)$ through its potential by $f=-\frac{dF}{dX}$, where $F$ is given by,
\begin{equation}
F=-\frac{A_d X}{2} (u_{max}-X)-\frac{A_d^2 u_{max}^2}{4 A_s} \sqrt{1+(1-\frac{2(-BA_0+E+A_s X)}{A_d u_{max}})^2}.
\end{equation}
Solid and dashed lines in Fig.  \ref{Fig5S} show that depending on the parameter $Y=A_0$, the depth of the potential minima can change. The two minima are symmetric at $A_{0c}=2.05$, which implies that at $A_0=A_{0c}$, the front is motionless. In the table of pairs  of stable solutions at different $A_0$, the critical value $A_{0c}$ is the one which satisfies the equation $F(X_{01},A_{0c})=F(X_{02},A_{0c})$. The analytic expression for  $A_{0c}$ is cumbersome, so we placed the formula for $A_{0c}$ in appendix E. 

\subsection{Heteroclinic connection}
\begin{figure}
\vspace{-0.in}
\includegraphics[width=0.7\linewidth,height=0.5\linewidth,scale=0.5]{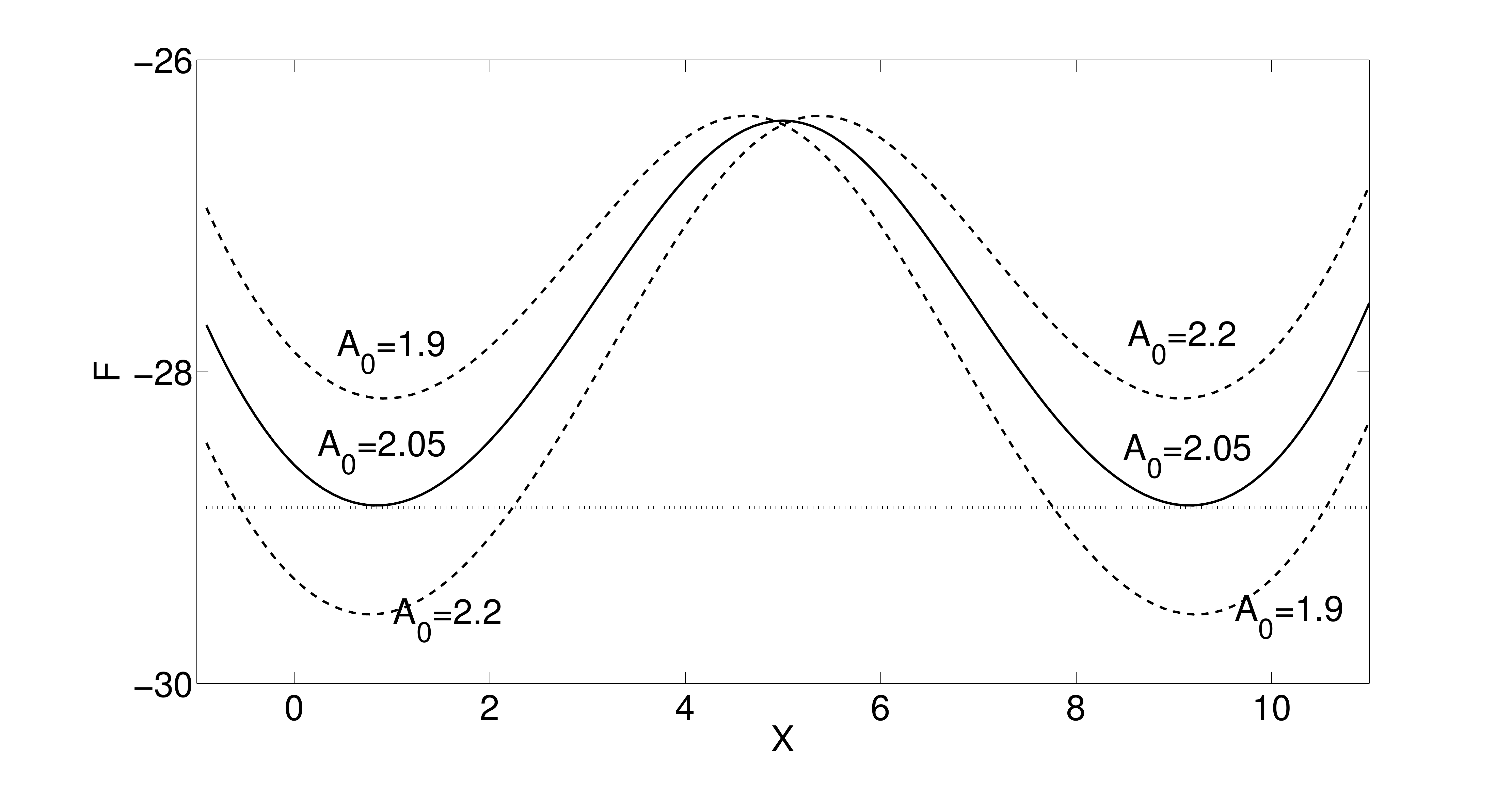} 
\caption{\label{Fig5S} The potential in Eq. (B2) at three different values of $A_0$.  When $A_{0}<A_{0c}=2.05$, the front moves in the direction of the expansion of $X_{02}$ solution. When $A_{0}>A_{0c}=2.05$, the front moves in the direction of the expansion of the $X_{01}$ solution.  At $A_0=A_{0c}$, the front is standing. Other parameters are the same as in Fig.  \ref{Fig4S}.}
\end{figure}

We confirmed the results shown in Fig.  \ref{Fig5S} via numerical bifurcation analysis. We transformed Eq. (B1) into two coupled ODE's, by introducing $z= x +vt$ and $X(x,t)=u(z)$,
\begin{eqnarray}
u_1'(z)=u_2(z), \nonumber\\
u_2'(z)=vu_2(z)-f(u_1(z),A_0),
\end{eqnarray}

\begin{figure}
\includegraphics[width=0.6\linewidth,height=0.4\linewidth,scale=0.5]{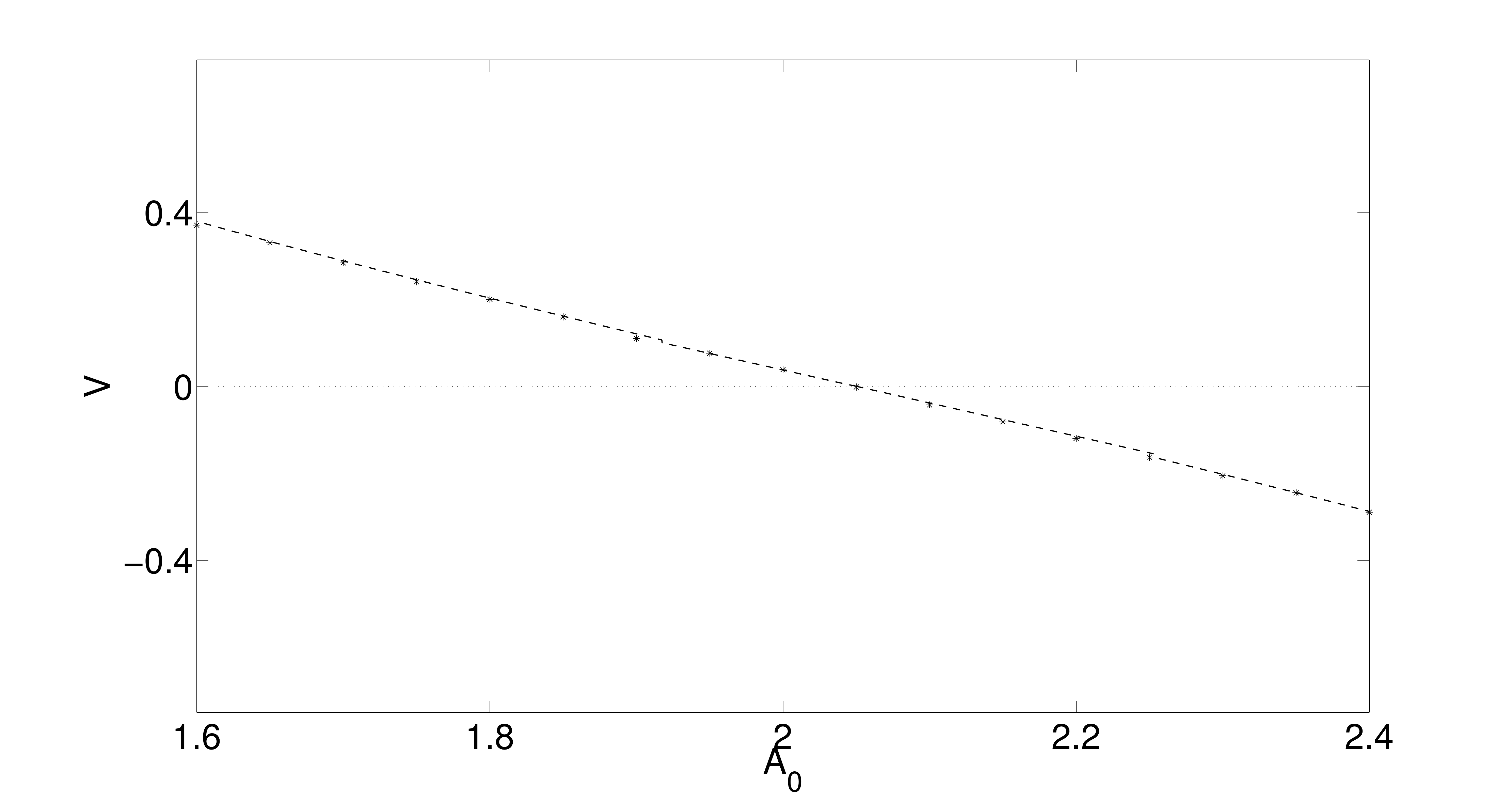} 
\caption{\label{Fig6S} Front velocity $v$ vs $A_0$. The dashed line is obtained via bifurcation analysis of Eq. (B3), as the heteroclinic connections of the steady state solutions. The symbols are obtained from the simulations of Eq. (B1). }
\end{figure}

where $\prime=\frac{d}{dz}$. Eq. (B3) has a pair of stationary solutions $(u_{10},u_{20})$ and $(u_{11},u_{21})$. The Jacobian of  Eq. (B3) is given by $J=(\{0,1\},\{\tilde B,v\})$, where $\tilde B=-(\frac{\partial{f(u_1(z),A_0)}}{{\partial u_1}})_{|\bf u_0}$. For the parameters shown in Fig.  \ref{Fig5S}, the pair of  solutions $(u_{10},u_{20})$ and $(u_{11},u_{21})$ are saddle points, as the Jacobian for these solutions have a pair of positive and negative eigenvalues. Using bifurcation analysis software AUTO-07p, we studied heteroclinic connections of $(u_{10},u_{20})$ and $(u_{11},u_{21})$, by using $A_0$ and $v$ as the bifurcation parameters. The bifurcation analysis is in agreement with Fig.  \ref{Fig5S} that at  $A_0=2.05$ the front is motionless. Also, the numerically computed velocities via bifurcation analysis and the front velocities computed from the simulations of Eq. (B1) are in  perfect agreement, Fig.  \ref{Fig6S}.

\begin{figure}
\vspace{-0.in}
\includegraphics[width=0.7\linewidth,height=0.5\linewidth,scale=0.5]{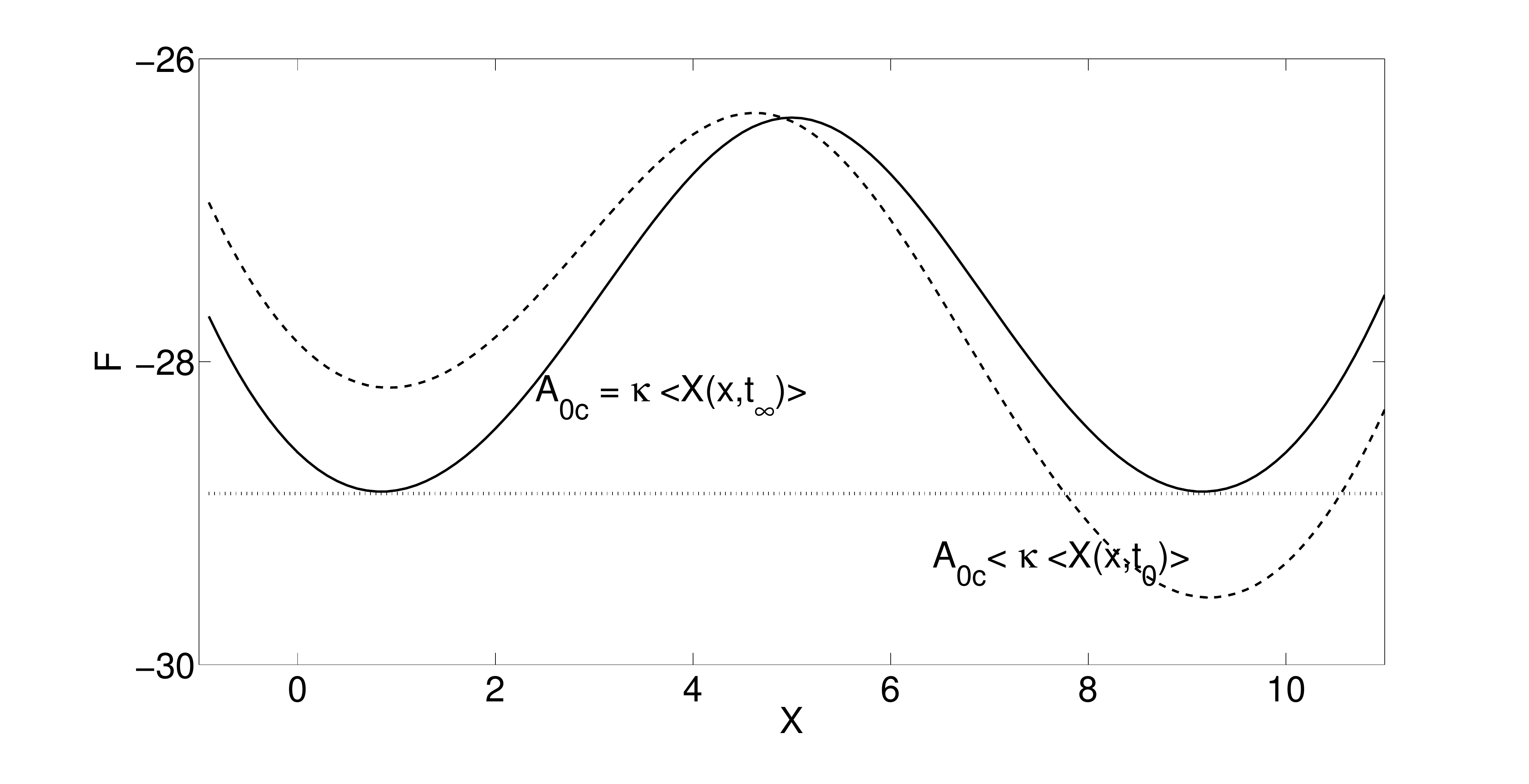} 
\caption{\label{Fig7S} The potential in Eq. (B2) in the case of global coupling. The parameters are the same as in Fig.  \ref{Fig5S}. When $\kappa \overline{X} < A_{0c}$, the front moves in the direction of the expansion of the $X_{02}$. Therefore, $\overline{X_{02}}$ increases with time, until the front reaches the point  $\kappa \overline{X} = A_{0c}$, where the front is motionless. }
\end{figure}

\subsection{Global Coupling}
Next we consider a case of $A_0= \kappa \overline{X}$, i. e. the global coupling case, where $\kappa$ is a constant. The PDE is now given by,
\begin{eqnarray}
\frac{\partial  X}{\partial t}=f(X) +\Delta X, \nonumber\\
f(X)=\frac{A_d u_{max}}{2} + \frac{{(E+A_s X-B \kappa \overline{X})-0.5 A_d u_{max}}}{{\sqrt{1+|\frac{2 (E+A_s X-B \kappa \overline{X})}{A_d u_{max}}-1|^2}}} -A_d X.
\end{eqnarray}

\begin{figure}
\includegraphics[width=0.6\linewidth,height=0.4\linewidth,scale=0.5]{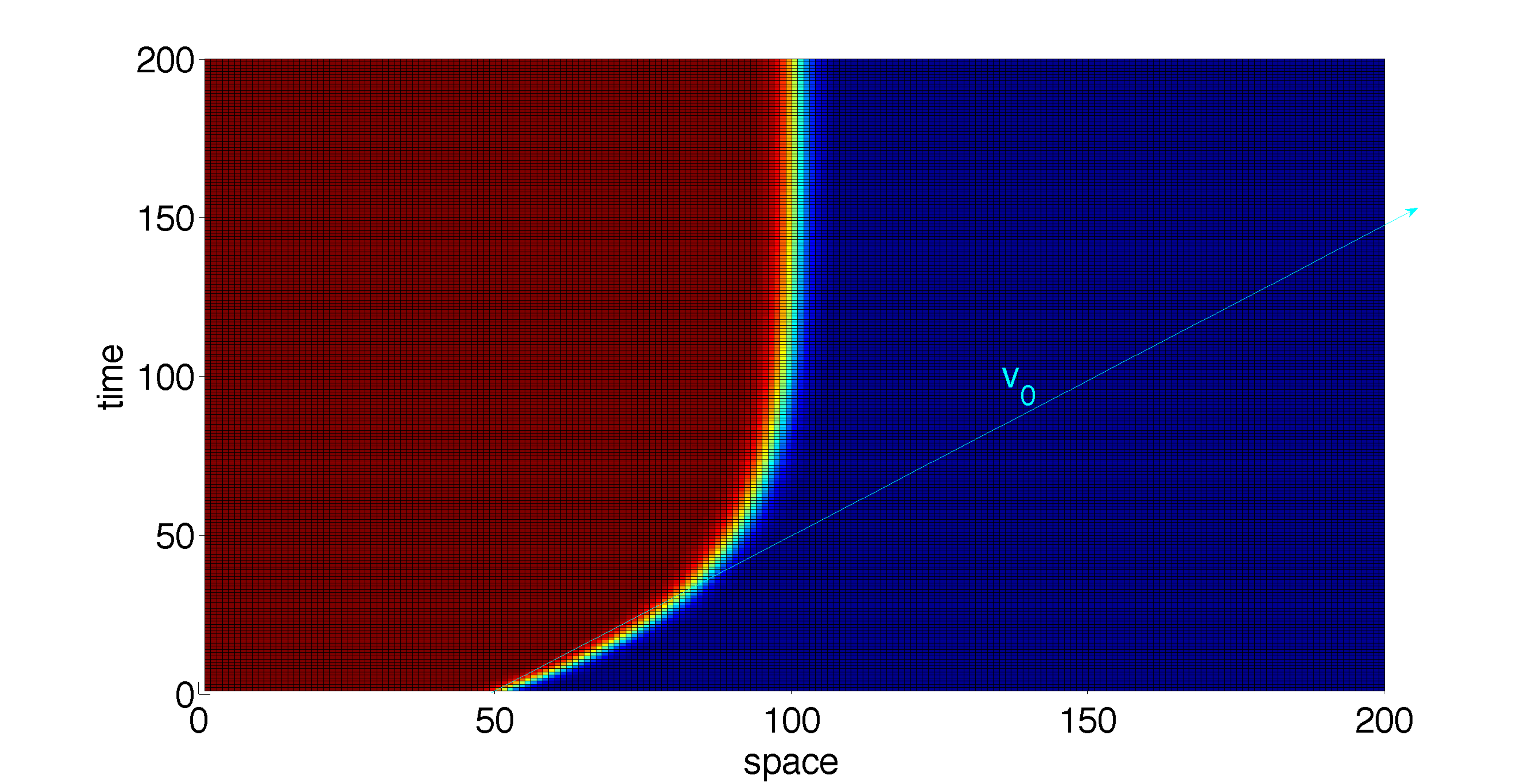} 
\caption{\label{Fig8S} (Color online) Domain confinement in the global coupling model, Eq.  (B4). In the case for $\kappa \overline{X(x,t)} = const$, the front propagates with velocity $v_0$. However, as $\kappa \overline{X(x,t)}$ increases, the front velocity decreases and eventually becomes zero at $\kappa \overline{X(x,t)} = A_{0c}$. No flux boundary conditions were used for the simulations of Eq. (B4) 
with the parameters in Fig.  \ref{Fig7S} , except $\kappa=0.16$.}
\end{figure}

When $\kappa$ and $\overline{X(x,t_0)}$ are small, i.e., $\kappa \overline{X(x,t)} < A_{0c}$,  the potential at the  steady state, $X_{02}$, has a deeper minimum, Fig.  \ref{Fig7S}. Therefore, the front will propagate in the direction  of expansion of the $X_{02}$ state Fig.  \ref{Fig8S},  and this process leads to the increase of $\overline{X(x,t)}$. However, the front propagation slows down and eventually stops as it approaches the point where $\kappa \overline{X(x,t)} = A_{0c}$. The final size of the $X_{02}$ domain, i.e. $\overline{X(x,t_{\infty})}$,  is controlled by the constant $\kappa$. If $\kappa$ is smaller, $\overline{X(x,t_{\infty})}$ is larger, and vice versa. 

\subsection{A two variable model}
Now we consider a two variable model for $A_0=S_Y H_0$,
\begin{eqnarray}
\frac{\partial  X}{\partial t}=\Phi(E+A_s X -B Y) -A_d X+ D_X \Delta X, \nonumber\\
\frac{\partial  Y}{\partial t}= C X -D Y +A_0 +D_Y \Delta Y.
\end{eqnarray}
Let us assume that $D_Y=0$ and there is a critical value $A_{0c}$ at which the front in the above equation is standing. Then we obtain $Y_0=\frac{C X +A_0}{D}$. By solving $f(X,A_0)=0$, with $\Phi(E+\tilde{A_s} X -B \tilde {A_0})$, where $\tilde{A_s} = A_s-\frac{ B C} {D}$, and $\tilde {A_0}=\frac{A_0}{D}$, we obtain the table of stationary solutions $X_{01}$ and $X_{02}$ at different values of $\tilde A_0$. The critical value of $\tilde A_{0c}$ and the corresponding stationary solutions $X_{01}$ and $X_{02}$ satisfy Eq. (E1). At the critical value of $\tilde A_{0c}$, the front is standing. 

If $D_Y\neq0$, Eq. (B5) is the continuous limit of the  two-variable Fujita et al. model for $A_0=0$. At the critical value of $A_0=A_{0c}$,  periodic domain patterns are possible when the ratio $\frac{D_Y}{D_X}\geq 1$. Moreover, if we assume $Y$ to be a fast variable, domain nucleation and domain confinement can be found in such a two-variable model.

\subsection{Domain Confinement in the Three Variable Model}
\begin{figure}
\includegraphics[width=0.6\linewidth,height=0.4\linewidth,scale=0.5]{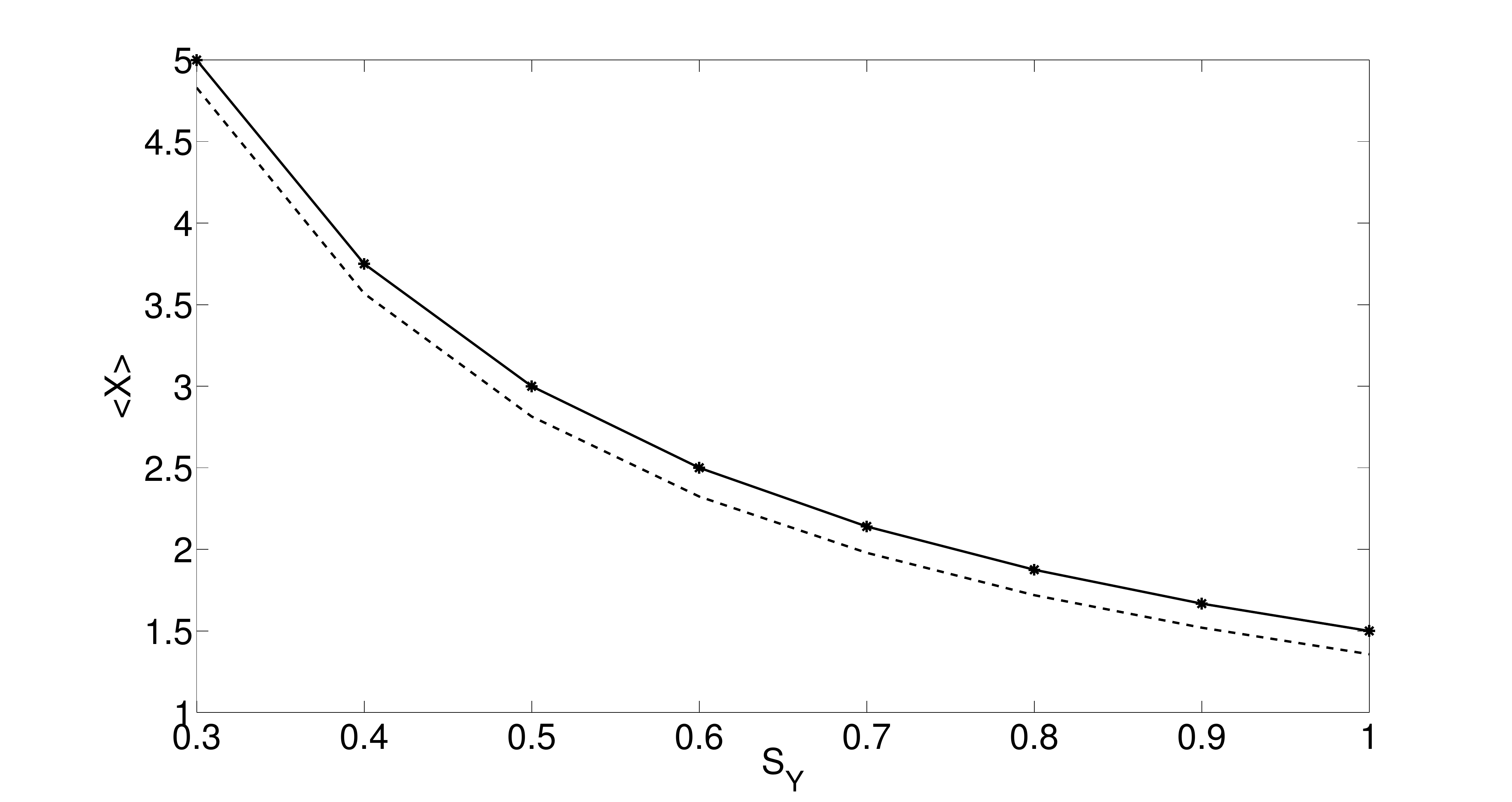} 
\caption{\label{Fig9S} Solid lines show domain confinement in  Eq.  (B5) when $A_0$ is replaced by $S_Y \overline{X(x,t)}$ and $D_Y=0$. Dashed lines show simulations of Eq. (B6). Other parameters are the same as in the Fig. 5 of the main text.}
\end{figure}

Finally, let us consider  the case of diffusive $X$ and $Y$ in the full model,
\begin{eqnarray}
\epsilon \frac{\partial  H}{\partial t}=-H + D_H \Delta_x H + X, \nonumber\\
\frac{\partial  X}{\partial t}=\Phi(E+A_s X -B Y) -A_d X+ D_X \Delta X, \nonumber\\
\frac{\partial  Y}{\partial t}= C X -D Y +S_Y H+D_Y \Delta Y.
\end{eqnarray}
It can be shown numerically that the same mechanism, based on the equilibrium of the domain potentials, as we have shown above is responsible for the domain confinement, when $\epsilon<<1$ and $D_H>>1$. The dotted lines in Fig.  \ref{Fig9S} show the domain confinement in the three  variable model, Eq. (B6), at different values of $S_Y$. The solid lines in Fig.  \ref{Fig9S} show the results of simulations of Eq. (B5), when $A_0$ is replaced by the global coupling term, $A_0=S_Y \overline{X(x,t)}$. The symbols show that the stationary distributions obtained from the simulations  of Eq. (B5) with different initial conditions fit well the relationship, $\overline{X}=\frac{A_0}{S_Y}$, for $A_0=1.5$. Note that at $A_0=1.5$, Eq. (B2) has two symmetric minima for the parameters in Fig. 5. According to Fig.  \ref{Fig9S}, the parameter $S_Y$ controls the size of the confined domain. Thus, the results in this section demonstrate that a nonuniform field $H$ can enforce a domain confinement.

\section{Wavenumber Instability}
Numerical simulations show that the homogenous steady states $(X_{01},Y_{01})$ and $(X_{02},Y_{02})$ can be unstable against small fluctuations, and complex patterns can be formed near the saddle-node points $SN_1$ and $SN_2$ in Fig. 1 of the main text. The wavenumber instability of the uniform solutions, ${\bf X_0}= (X_{02},Y_{02})$ (or ${\bf X_0}= (X_{01},Y_{01})$),  can be analyzed  by putting the perturbed solutions, 
${\bf X}= {\bf X_0}+ \mathrm{e}^{\lambda t} \mathrm{cos(q x) }\delta{\bf X}$,  into 
 \begin{eqnarray}
\frac{\partial  X}{\partial t}=\Phi(E+A_s X -B Y) -A_d X +D_X \Delta X, \nonumber\\
\frac{\partial  Y}{\partial t}= C X -D Y +HS_0+{\frac{S_Y}{2\sqrt{D_H}}} \int_0^L \mathrm{e}^{-\frac{ |x-x'|}{\sqrt{D_H}}}X(x',t)\,\mathrm{d}x' +D_Y \Delta Y.
\end{eqnarray}
After standard calculations, the characteristic equation for the stability of the  uniform-state is given by,
\begin{equation}
(\lambda -\tilde{\Phi_x}+A_d+D_X q^2)(\lambda+D+D_Y q^2)=\tilde{\Phi_y} (C'+K_y),
\end{equation}
where, $K_y=-S_Y \frac{q^2}{\kappa'^2+q^2}$, $\kappa'=\sqrt{D_H^{-1}}$, $C'=C+S_Y$, $\tilde {\Phi_x}={\frac{As}{{(O^n+1)}^{\frac{n+1}{n}}}}$, $\tilde {\Phi_y}={\frac{-B}{{(O^n+1)}^{\frac{n+1}{n}}}}$, and $O=1+{(\frac{2 (E+A_s X_0 -B Y_0) -A_d u_{max}}{A_d u_{max}}})^n$.

\begin{figure}
\vspace{-0.in}
\includegraphics[width=0.7\linewidth,height=0.4\linewidth,scale=0.3]{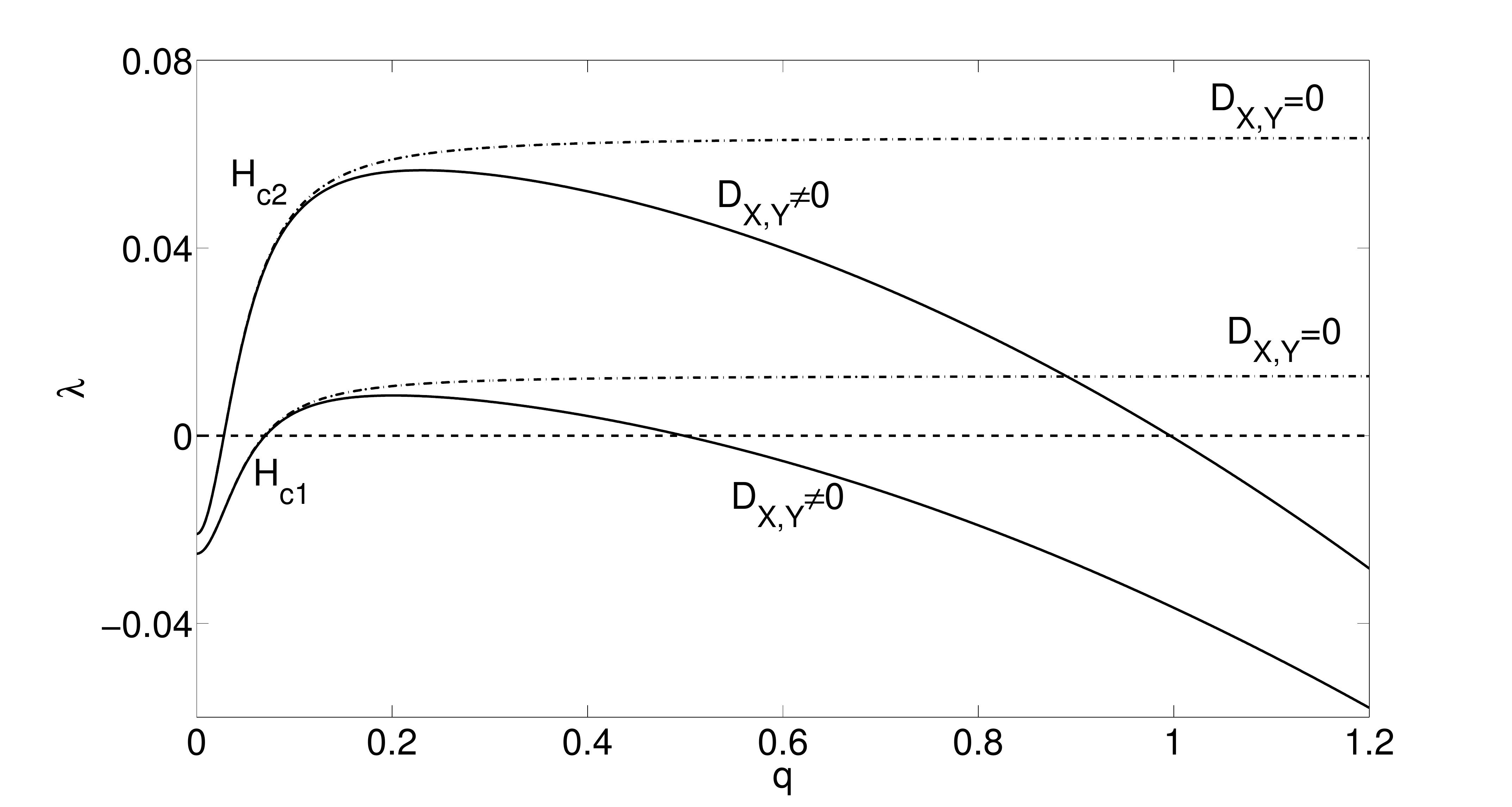} 
\caption{\label{Fig10S} Linear stability spectra of the uniform states near saddle-node points. Parameters are: $H_{c1}$ : $HS_0=0.442$ and $S_Y=0.055$;   $H_{c2}$ : $HS_0=1$ and $S_Y=0.0455$  . Other parameters are the same as in Fig. 6 of the main text and $D_X=0.1$, $D_Y=0.1$, and $\kappa'=0.05$. }
\end{figure}

In Fig.  \ref{Fig10S} we show the spectra of $\lambda$ for the uniform states $(X_{01},Y_{01})$ near the point $SN_1$ and $(X'_{02},Y'_{02})$ near the point $SN_2$. Note that the corresponding upper and lower uniform states, $(X_{02},Y_{02})$ and $(X'_{01},Y'_{01})$, are stable against small nonuniform perturbations. When $D_X=D_Y=0$, the spectra are positive,  $\lambda\geq 0$, at all $q>q_{c}$, meaning that the neighboring elements can have sharp discontinuities.  The two features, $\lambda\leq 0$ at $q<<1$ and $\lambda\geq 0$ at $q>>1$ imply that sharp discontinuities and extended uniform-like states may coexist in the system. It explains the observation of the {\em chimera} like  states in the simulations. When $D_X \neq 0$ and $D_Y \neq 0$, the spectra are negative, $\lambda\leq 0$, at large wavenumbers  $q>>1$, meaning that there are no sharp discontinuities of the state variables between the neighboring elements. 

The linear stabiliy analyses in Fig.  \ref{Fig10S} are  in agreement with simulations. In the simulations shown in Fig. 6 of the main text, a wavenumber instability does not lead to stationary periodic patterns near the upper steady state; instead, it may generate states near the lower steady state, if the local values of $H(x,t)$ exceed $HS_{c2}$. When $D_X$ and $D_Y$ are small, the generated stationary patterns often form irregular domains.

\section{Comparison with Turing Patterns in monostable systems}
\begin{figure}
\includegraphics[width=1\linewidth,height=0.4\linewidth,scale=0.5]{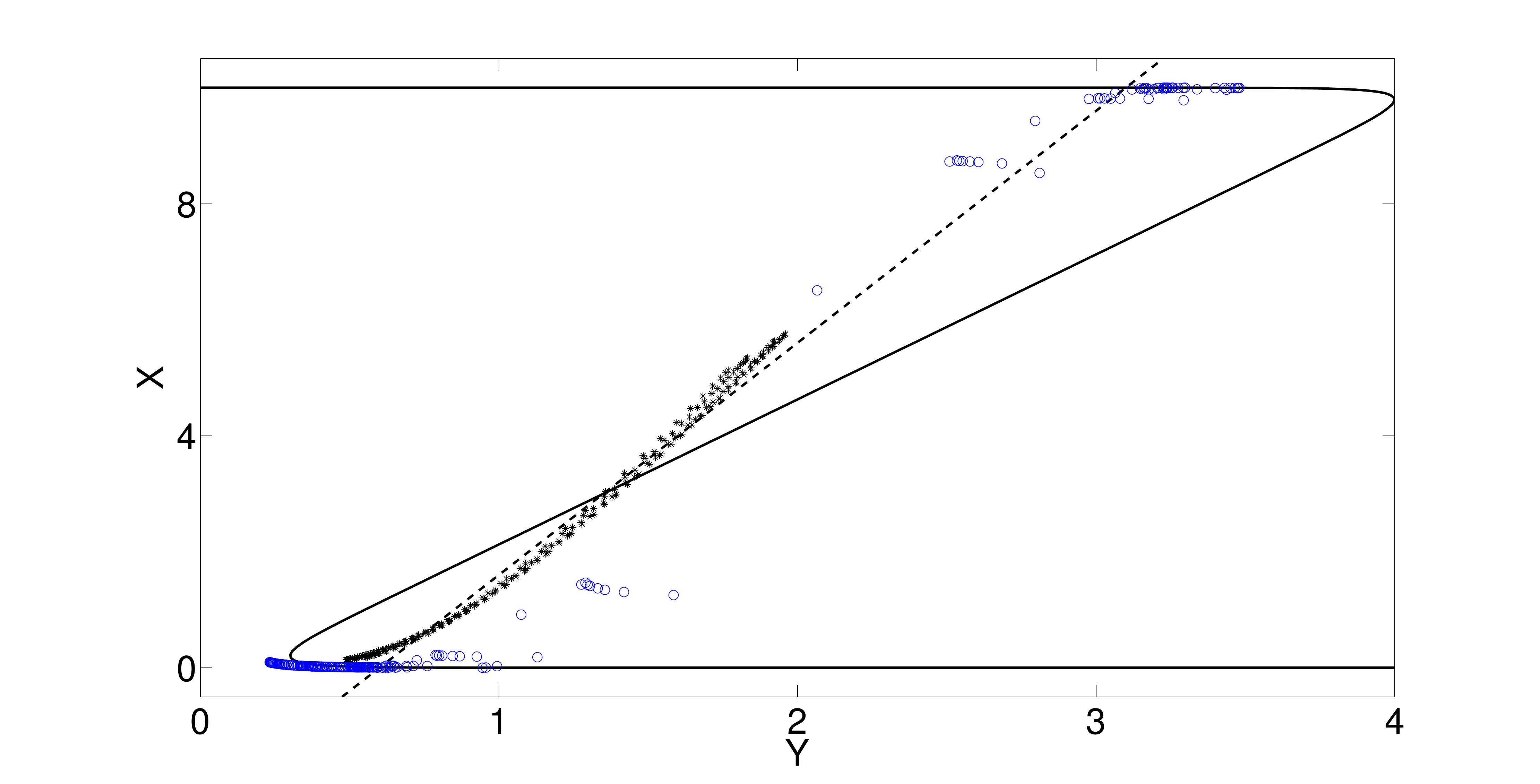} 
\caption{\label{Fig12aS} (Color online) Projection of patterns on $(Y,X)$ plane.  Black symbols: Turing patterns near the saddle-node point $HS_1$ in Fig. 1 Parameters are: $E=0.3$, $D_H=0.36$,   and $\delta x=0.05$. Blue symbols(open circles): patterns in the bistable region. Parameters are: $E=0.1$, $D_H=100$,  and $\delta x=0.5$.  Other parameters are: $B=2$, $C=0.25$, $D=1$, $A_d=1$, $A_s=1.8$, $u_{max}=10$, $D_X=0.05$, $D_Y=0.1$, $S_Y=0.25$,  $N=256$, $\epsilon=1$, and  $n=20$. }
\end{figure}

\begin{figure}
\includegraphics[width=1\linewidth,height=0.4\linewidth,scale=0.5]{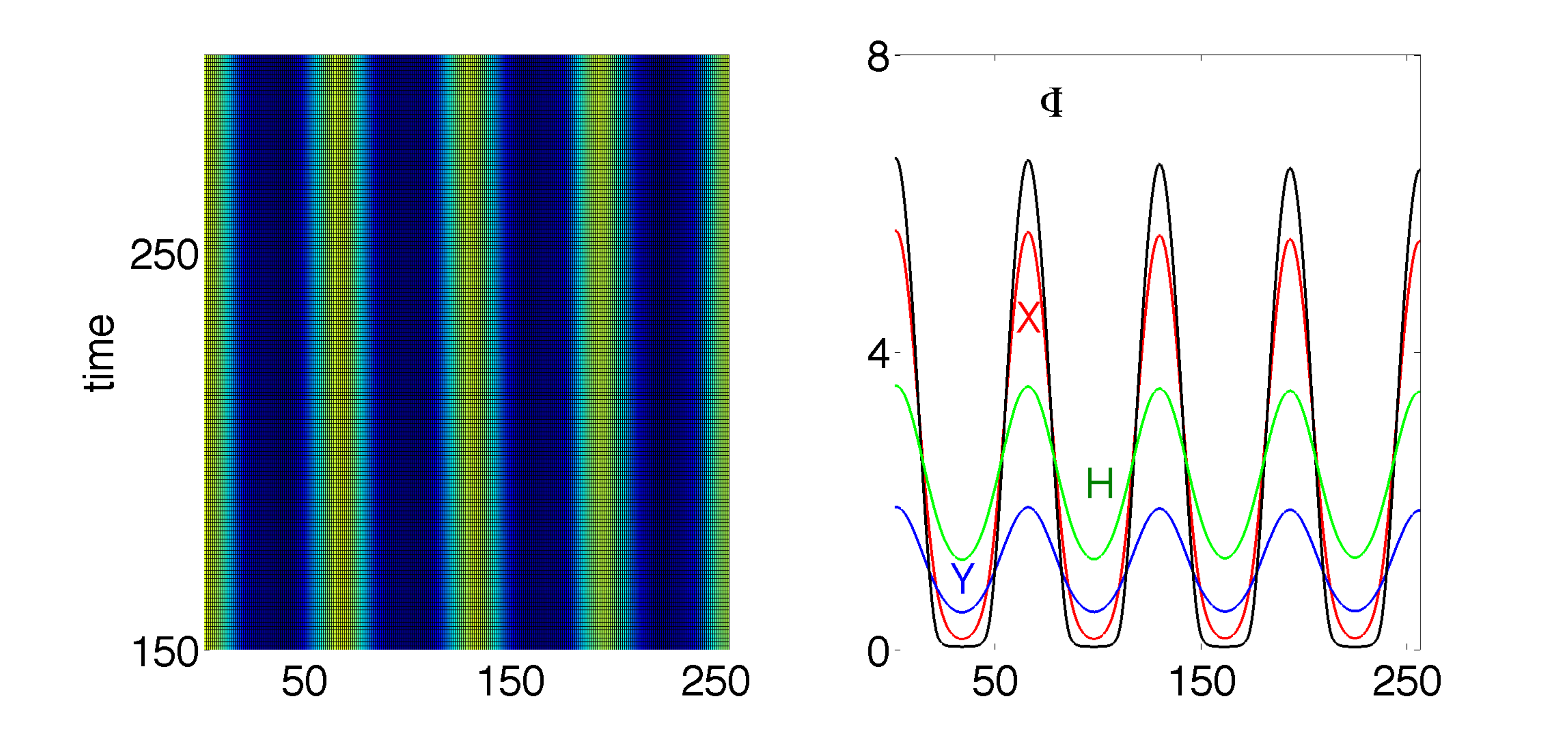} 
\caption{\label{Figbrus1} (Color online) Space-time profiles of the stationary Turing  patterns shown by black symbols in Fig. \ref{Fig12aS}. Formation of the pattern does not involve the mechanism described in Fig. 2. }
\end{figure}

\begin{figure}
\includegraphics[width=1\linewidth,height=0.4\linewidth,scale=0.5]{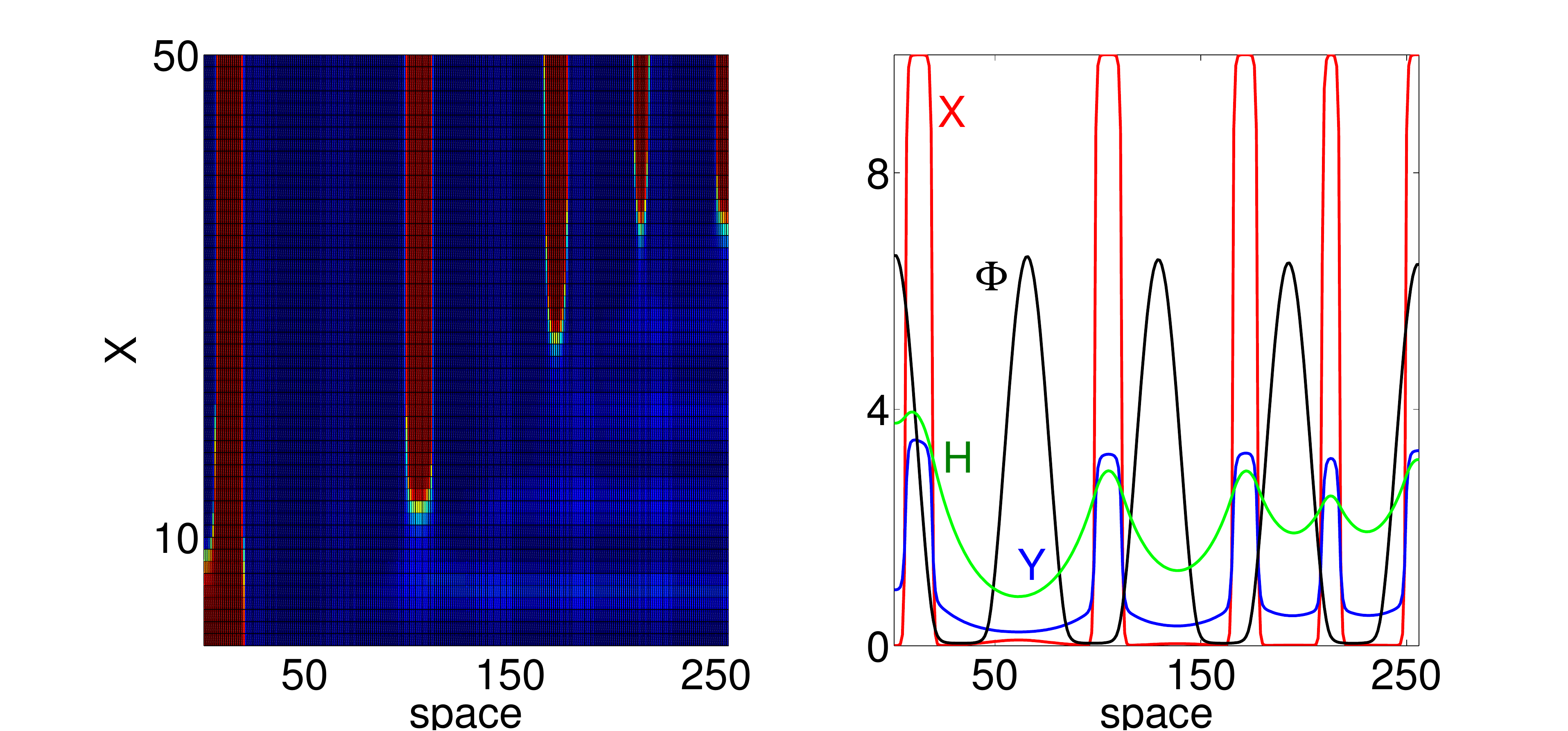} 
\caption{\label{Figbrus2} (Color online) Space-time profiles of the stationary patterns shown by blue symbols(open circles)  in Fig. \ref{Fig12aS}. Formation of the pattern does involve the mechanism described in Fig. 2. }
\end{figure}

When $D_H \sim D_{X,Y}$,  stationary Turing patterns are possible in Eqs. (2-3). On the $(Y,X)$ plane these regular patterns populate the space in between the bistable states, black symbols in Fig. \ref{Fig12aS}, whereas the patterns that develop for  $D_H>>1$ connect the bistable states, blue symbols (open circles)  Fig. \ref{Fig12aS}. The black symbols represent regular Turing patterns, with a selected wavenumber, Fig.  \ref{Figbrus1}. The blue symbols (open circles)  can be considered as a continuation of the Turing patterns into the bistable region, emerged from a homogeneous $(X_{01},Y_{01})$ state near the saddle-node bifurcation point,  Fig.  \ref{Figbrus2}. However, in contrast to Turing patterns in monostable systems, there is no clear maximum in the Fourier spectrum of the patterns in the bistable region, where the patterns can become highly  irregular due to the interplay with bistability. The size, location and spatiotemporal dynamics of irregular patterns can be explained by the domain confinement mechanism we described above.  

\begin{figure}
\includegraphics[width=1\linewidth,height=0.4\linewidth,scale=0.5]{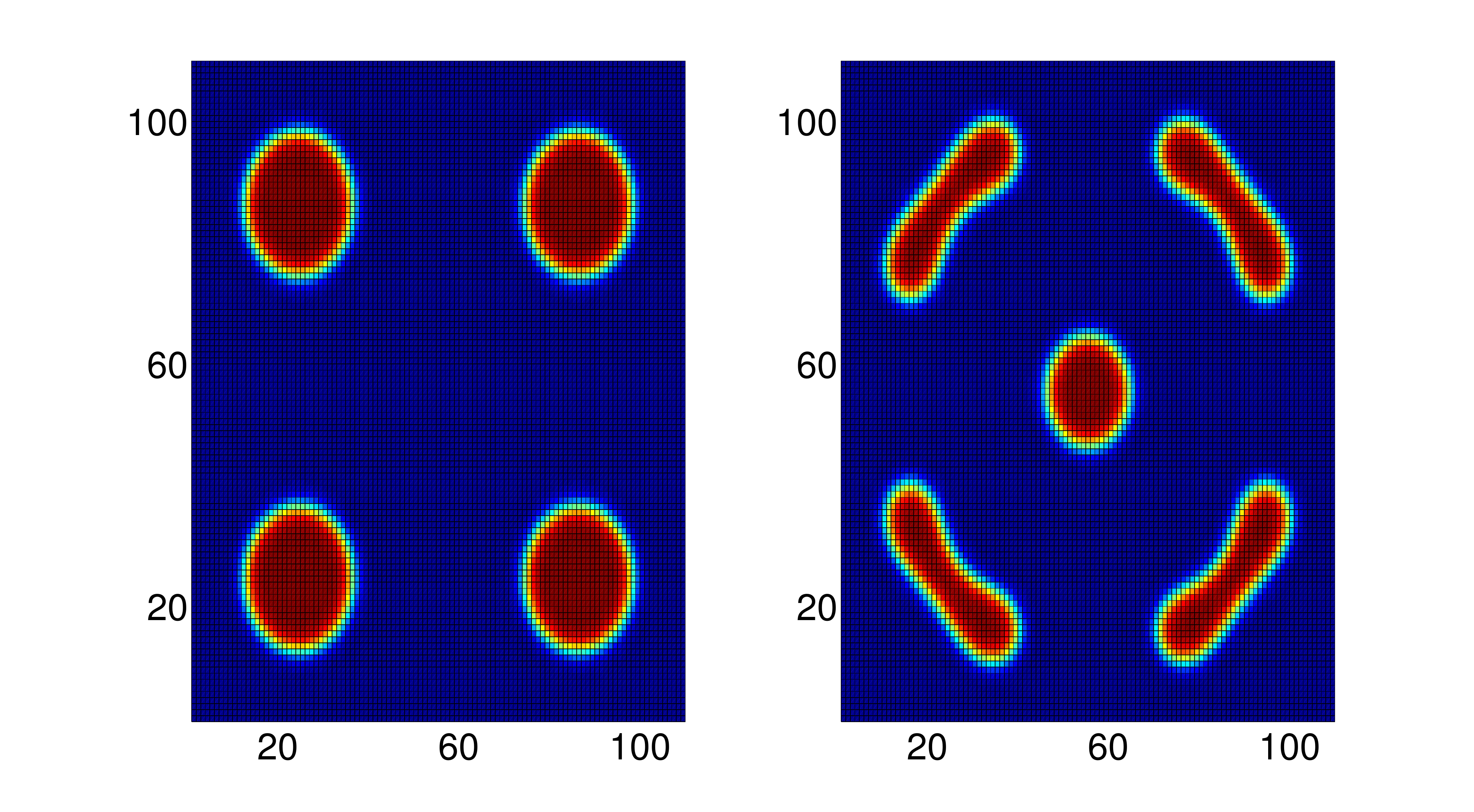} 
\caption{\label{Fig11S} (Color online) $X$ patterns in space dimension two  for simulations with no-flux boundary conditions, from the same initial conditions.  Parameters are: $E=1$, $B=2.5$, $C=0.25$, $D=1$, $A_d=1$, $A_s=1.9$, $u_{max}=10$, $D_X=0.3$, $D_Y=0.3$, $S_Y=0.5$, $D_H=100$, and  $n=20$. Left: $\epsilon=0$, Turing like patterns.  Right: $\epsilon=0.01$, irregular patterns.}
\end{figure}

Fig.  \ref{Fig11S} shows two dimensional stationary patterns obtained from simulations with no-flux boundary conditions. The plot on the left resembles Turing patterns, while the plot on the right looks different than typical Turing patterns near the saddle-node points; for example, hexagons or regular spots. These two plots, selected as an example,  show that our model can display patterns both similar and dissimilar to the Turing patterns in reaction diffusion systems with monostable states.

\section{Calculation of $A_{0c}$}
The formula for $A_{0c}$ is given by
\begin{equation}
A_{0c}=\frac{1}{2B}(2 E + \frac{1}{u_{max}^2}(\alpha \pm  \beta)), 
\end{equation}
where,
\begin{widetext}
\begin{equation}
\alpha=A_s X_{01}^3 +A_s X_{01}^2 X_{02}-2A_s X_{01}^2 u_{max}-A_s X_{01} X_{02}^2+2 A_s X_{01} u_{max}^2 -A_s X_{02}^3 +2A_s X_{02}^2 u_{max} -A_d u_{max}^3.
\end{equation}
\end{widetext}
In Eq. (E1) $\beta$ is given by,
\begin{widetext}
\begin{equation}
\beta=(X_{01}+X_{02} -u_{max})(-A_s X_{01}^2 +A_s u_{max} X_{01} +A_s X_{02}^2 - A_s u_{max} X_{02} +\gamma),
\end{equation}
\end{widetext}
where, $\gamma$ is given by
\begin{widetext}
\begin{eqnarray}
\gamma=\frac{u_{max}}{(X_{01}+X_{02} )(X_{01}+X_{02} -2 u_{max})}\sqrt{\gamma_1+\gamma_2},\\
\gamma_1=-A_d^2 u_{max}^4 +A_s^2  X_{01}^4 -2 A_s X_{01}^3 u_{max} -2 A_s^2  X_{01}^2  X_{02}^2, \\
\gamma_2=2 A_s^2  X_{01}^2  X_{02} u_{max} +2 A_s^2  X_{01}  X_{02}^2 u_{max} +A_s^2 X_{02}^4 -2 A_s^2 X_{02}^3 u_{max},
\end{eqnarray}
\end{widetext}

\begin{acknowledgments}
I am thankful to Professors A. S. Mikhailov and  J. J. Tyson  for fruitful discussions. This work has been partially supported by the grant from Mongolian Science and Technology Foundation awarded to Prof. Kh. Namsrai.
\end{acknowledgments}

\bibliography{basename of .bib file}

\end{document}